\documentclass[mreferee,extra]{gji}
\usepackage{timet}
\usepackage{color}
\usepackage{floatflt}
\usepackage{verbatim}
\usepackage{graphicx}
\usepackage{epsfig}
\usepackage{subfigure}
\usepackage{amsmath}
\usepackage[normalem]{ulem}
\usepackage[urlcolor=blue,citecolor=black,linkcolor=black]{hyperref}
\usepackage{mathtools}

\newcommand{\wb}{\mathbf{w}}
\newcommand{\vb}{\mathbf{v}}
\newcommand{\ub}{\mathbf{u}}

\newcommand{\x}{\mathbf{x}}
\newcommand{\xr}{\mathbf{x}_r}
\newcommand{\xs}{\mathbf{x}_s}
\newcommand{\Gb}{\mathbf{G}}
\newcommand{\Nr}{\mathbf{N}_r}
\newcommand{\Kb}{\mathbf{K}}
\newcommand{\nb}{\mathbf{n}}
\newcommand{\Lb}{\mathbf{L}}
\newcommand{\Lbdag}{\mathbf{L}^{\dagger}}
\newcommand{\s}{\mathbf{s}}
\newcommand{\Wr}{\mitbf{\Lambda}}
\newcommand{\Wradj}{{\mitbf{\Lambda}}^{\dagger}}
\newcommand{\zero}{\mathbf{0}}
\newcommand{\Fm}{\mathcal{F}}
\newcommand{\Fb}{\mathbf{F}}
\newcommand{\Mm}{\mathcal{M}}
\newcommand{\Dm}{\mathcal{D}}
\newcommand{\lan}{\left\langle}
\newcommand{\ran}{\right\rangle}

\title[Target-enclosing inversion using an  interferometric objective function]
  {Target-enclosing inversion using an  interferometric objective function}
\author[P. Zheglova et al.]
  {Polina Zheglova$^{1,3}$, Matteo Ravasi$^1$, Ivan Vasconcelos$^2$, Alison Malcolm$^3$\\
  $^1$ King Abdullah University of Science and Technology, Thuwal 23955-6900, Saudi Arabia \\
  $^2$Utrecht University, Utrecht, Netherlands \\
  $^3$Memorial University of Newfoundland, St. John's, Canada
  }
\date{Received 2021 November ..; in original form 2021 November ..}
\pagerange{\pageref{firstpage}--\pageref{lastpage}}
\volume{200}
\pubyear{1998}

\begin{document}

\label{firstpage}

\maketitle

\begin{summary}
Full waveform inversion is a high-resolution subsurface imaging technique, in which full seismic waveforms are used to infer subsurface physical properties. We present a novel, target-enclosing, full-waveform inversion framework based on an interferometric objective function. This objective function exploits the equivalence between the convolution and correlation representation formulas, using data from a closed boundary around the target area of interest. Because such equivalence is violated when the knowledge of the enclosed medium is incorrect, we propose to minimize the mismatch between the wavefields independently reconstructed by the two representation formulas. The proposed method requires only kinematic knowledge of the subsurface model, specifically the overburden for redatuming, and does not require prior knowledge of the model below the target area. In this sense it is truly local: sensitive only to the medium parameters within the chosen target, with no assumptions about the medium or scattering regime outside the target. We present the theoretical framework and derive the gradient of the new objective function via the adjoint-state method and apply it to a synthetic example with exactly redatumed wavefields. A comparison with  FWI of surface data and target-oriented FWI based on the convolution representation theorem only shows the superiority of our method both in terms of the quality of target recovery and reduction in computational cost. 
\end{summary}


\section{Introduction}

Reflection seismology plays a key role in the discovery and management of underground resources ranging from coal to hydrocarbons. As society progresses towards interacting with the Earth's subsurface in a more sustainable manner via geothermal production, carbon and hydrogen capture and storage, and hydrogen production, the impact of seismic data in the decision making process is likely to continue to be vital. To this end, target-oriented waveform inversion is becoming increasingly popular under the promise of significantly reducing the computational cost of estimating high-resolution models for specific areas of interest in the subsurface. This is particularly appealing in the context of reservoir characterization and time-lapse inversion -- which are key to sustainable subsurface management --  where the overall kinematics of the model is usually already known from baseline data. 

Three alternative approaches to target-oriented waveform inversion have emerged in the literature: the first relies on so-called local solvers to compute the wavefields in the region of interest to be used to update the model parameters \citep{Malcolm:2016, Willemsen:2016, Willemsen:2017, Kumar:2019, Jaimes-Osorio:2020, Jaimes-Osorio:2021}; the second leverages model-based \citep{Costa:2018, Garg:2020} or data-driven \citep{Cui:2020} receiver-side redatuming to produce up- and down-going separated wavefields at depth. Such wavefields are subsequently used as input to an objective function based on the convolutional-type representation theorem. In other words, the upgoing wavefield is modelled by means of multi-dimensional convolution between the available downgoing wavefield and the reconstructed local reflection response, which is a function of the model parameters in the target area of interest. The objective function is minimized by improving the model parameters such that the modelled upgoing field approaches the one previously retrieved by means of redatuming. A third approach \citep{Yang:2012, Biondi:2018, Guo:2020, Biondi:2020, Li:2021} relies on the ability to redatum both sources and receivers to the target level of interest, rendering a local reflection response that is, in principle, completely independent of the overburden. Such virtual reflection data is subsequently simply used as input to a conventional surface FWI engine. A main drawback of model-based approaches lies in the fact that the quality of the the source- and receiver-side wavefield propagators relies heavily on the accuracy of a first step of velocity-model building and its ability to retrieve high-frequency components of the subsurface that allow for inclusion of multiply scattered waves in the propagators. On the other hand, data-driven approaches that rely for example on Marchenko redatuming  \citep[e.g.,][]{Neut:2015} only require a kinematically accurate, smooth model of the overburden. Moreover, despite early successful results, all of the aforementioned methods apart from that of \cite{Cui:2020} are not fully target-oriented as they require estimating model parameters, to a relatively high degree of accuracy, in an infinite half-space from the datum of interest in the subsurface.

We propose a new full waveform inversion (FWI) framework to invert for the velocity model locally from full waveforms redatumed to an enclosing boundary of the subsurface subdomain of interest based on the objective function initially proposed by \cite{Vasconcelos:2016}. This objective function is conceptually different from both the traditional FWI misfit function and those used in previous approaches to target-oriented inversion: instead of relying on the equivalence of the source and receiver wavefields at the physical (or virtual) receiver locations, it  builds on the equivalence of the wavefields reconstructed from the boundary data by the convolution and correlation representation formulas anywhere in the local subdomain. Examples of such representation formulas can be found, in e.g. classical texts such as \cite{Morse:1953} (acoustic) and \cite{Aki:2002} (elastic).

The correlation-type representation formulas are widely applicable in geophysics, e.g. in seismic interferometry \citep{ Curtis:2006, Wapenaar:2006}, backward wavefield extrapolation \citep{Vasconcelos:2013, Ravasi:2015}, and non-linear imaging  \citep{Fleury:2012, Ravasi:2014}. In addition, the correlation-type representation formulas find application in theoretical time-reversal acoustics \citep{Cassereau:1992, Fink:2001}. Simiarly, the convolution-type representation formulas are used in the geophysical context for seismic interferometry \citep{Wapenaar:2011}, forward wavefield extrapolation \citep{Robertsson:2000, Vasmel:2016}, immersive boundary conditions \citep{vanManen:2007}, and the local domain methods cited in the previous paragraph. Non-geophysical applications include, e.g., sound wavefield synthesis \citep{Spors:2008}. \cite{Wapenaar:2007} presents a general formulation of the convolution and correlation representation theorems for a class of wave propagation problems arising in geophysics. 

The idea of the proposed method stems from the fact that in a lossless medium, both the convolution and correlation representation formulas can be used to reconstruct the true wavefields in an open subdomain from wavefield measurements on the subdomain boundary, if the Green's function, i.e. model parameters within the subdomain are exactly known. Moreover, the wavefields reconstructed by the convolution and correlation representation theorems are the same up to the time direction. If the Green's function, i.e. the model parameters are incorrectly known, then these fields have similar features but differ in details. Our inversion scheme is therefore driven by the mismatch of the wavefields reconstructed by the convolution and correlation representation formulas when the model is incorrect. In the ideal situation,  when the boundary data are exact, the mismatch between the convolution and correlation wavefields is nil when all the features of the true model are recovered by the inversion. Thus, the proposed method represents an ideal platform for local refinement of the subsurface model parameter distribution. 

Moreover, as the boundary data contain all of the waves entering, leaving and re-entering the local domain, all of these wave phases are utilized by the inversion. Also, since the convolution and correlation representation formulas reconstruct the wavefields only within the injection boundary, the method is insensitive to model inaccuracies outside of the injection surface. Therefore, the proposed inversion method does not require us to invert for an infinite half-space below the top boundary. The modelled wavefields include the waves entering the local subdomain from below without the necessity to model wave propagation in the underburden at each step of the inversion. In other words, the local domain forward modelling process accounts for all events coming from outside of the local subdomain. 

The proposed method requires the knowledge of the wavefields on a closed boundary completely surrounding the local subdomain. In practice such data can either be measured by borehole receivers located underground, or, more often, they are obtained by redatuming methods, e.g., Marchenko redatuming \citep{Wapenaar:2014, Neut:2015, Ravasi:2017, Vargas:2021}. Within the context of local waveform inversion based on redatumed wavefields and the convolution representation formula, \cite{Cui:2020} quantitatively assess the retrieval accuracy of the Marchenko method. They show that inversion with redatumed boundary data is almost as accurate as the inversion with the exact boundary data, although, admittedly, they use a starting velocity model of a very high quality both for redatuming and local inversion. They also show that a local refinement of the subsurface model with the redatumed wavefields is comparable in resolution to the refinement obtained by full waveform inversion of surface data.

The main contribution of this work is to formulate a new full waveform inversion method based on the interferometric  objective function of \cite{Vasconcelos:2016} and investigate its properties. We formulate the interferometric objective function and the partial differential equation (PDE) constraints stemming from the convolution and correlation representation integrals, using the constant-density vector-acoustic formulation for pressure and particle displacement \citep{Fleury:2013,Zheglova:2020}, where the inverted model parameter is squared slowness. We derive the gradient of the interferometric objective function with respect to the model parameter by the adjoint state method. Then we implement an interferometric full waveform inversion (IFWI) method, where the model updates are produced with the help of an L-BFGS optimization engine. We investigate the properties of the interferometric objective function, its gradient and the new inversion method on a series of stylized synthetic examples, using exact (i.e., directly modelled) boundary wavefields. This choice allows us to focus on the features and effectiveness of the proposed approach in the ideal situation. We show that with exact data, the proposed method is able to better recover the low wavenumber components in the model compared to the conventional surface FWI method and the local FWI method based on the convolution representation formula \citep{Cui:2020}. Then we estimate with proxy examples the potential influence of using redatumed data. In particular, we investigate the influence on the inversion performance of kinematically incorrect boundary data, and the influence of missing data on the inversion result. Finally, we discuss potential outcomes and issues related to the implementation of the method with redatumed wavefields.

\section{The Method}

\subsection{Preliminaries}

In this section we review the vector-acoustic convolution and correlation representation formulas that will be used in the formulation of the interferometric full waveform inversion (IFWI) objective function. As shown by \cite{Wapenaar:2007}, equations (63) and (67), the vector-acoustic field $\wb = [p, \vb]$, where $p$ is pressure and $\vb$ is particle velocity, in an open subdomain $D$ of the subsurface can be exactly reconstructed in forward or reverse time order from measurements on the boundary $\partial D$, if the Green's function $\Gb_D$ in $D$ is known exactly, and the physical source of the wavefield is located outside of the subdomain $D$. Denoting the field reconstructed forward in time as the \textit{convolution} field, $\wb^{conv}$, and the field reconstructed in reverse time order as the \textit{correlation} field, $\wb^{corr}$, Wapenaar's equations are written in the frequency domain as:
\begin{eqnarray}
	\wb^{conv}(\x,\xs,\omega) &=& -\int_{\partial D} \Gb_D(\x,\xr,\omega) \Nr \wb(\xr,\xs,\omega) dS_r \label{eq:conv_rep}\\
	\wb^{corr}(\x,\xs,\omega) &=& \int_{\partial D} \Kb \Gb_D^{\ast}(\x,\xr,\omega) \Kb \Nr \wb(\xr,\xs,\omega) dS_r, \label{eq:corr_rep}
\end{eqnarray}
where 
\begin{description}
	\item $\x$ is a point in $D$;
	\item $\xr$ is a point on the boundary $\partial D$;
	\item $\xs$ is physical source of the field, located outside of $D \cup \partial D$;
	\item $\omega$ is angular frequency;
	\item $\Nr$ is the matrix containing the outward normal components $\nb$ to $\partial D$:
		\begin{equation}
			\Nr = \begin{pmatrix}
				0 & \nb^T\\
				\nb  & \mathbf{0}
			\end{pmatrix};
		\end{equation}
	\item $\Kb$ is the diagonal matrix encoding the change of sign in $\vb$ due to time-reversal:
		\begin{equation}
			\Kb = \begin{pmatrix}
				1 & \mathbf{0}^T\\
				\mathbf{0}  & -\mathbf{I}
			\end{pmatrix}.
		\end{equation}
\end{description}
Here, $\ast$ is the complex conjugation in the frequency domain corresponding to time-reversal in the time domain,  whilst $\mathbf{0}$ and $\mathbf{I}$ are the zero and identity matrices, respectively.

\begin{figure}\centering
 	\subfigure[]{\includegraphics[width = 0.3\textwidth,trim={0cm 2cm 0cm 0cm},clip]{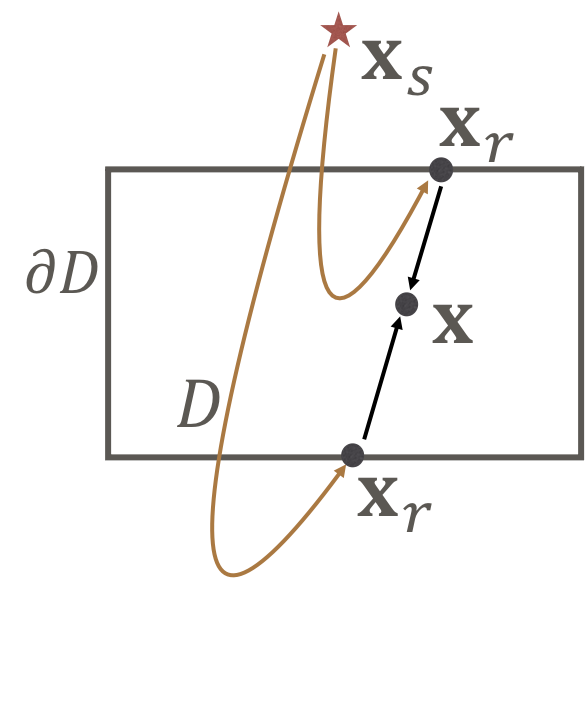}} \hspace{0.5cm}
	\subfigure[]{\includegraphics[width = 0.3\textwidth,trim={0cm 2cm 0cm 0cm},clip]{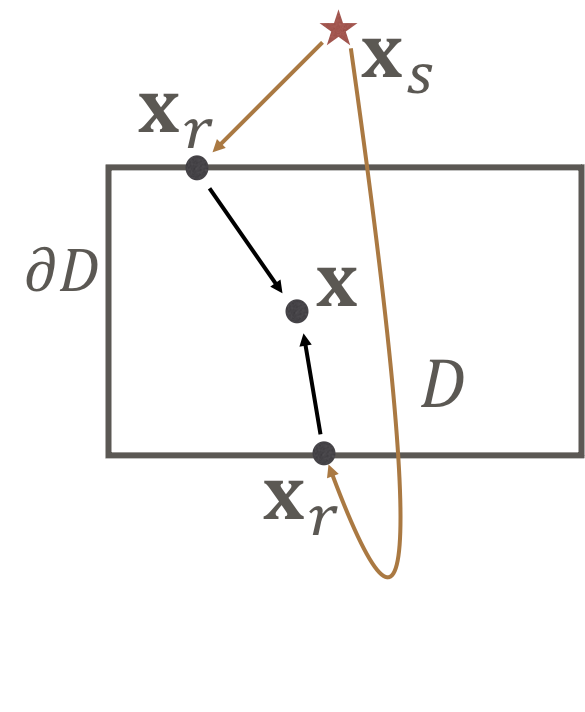}}
 	\caption{Schematic of wavefield extrapolation: (a) convolution, (b) correlation. The wavefield from the source $\xs$ is propagated to a point $\xr \in \partial D$, from which it is reconstructed at $\x \in D$ using equations (\ref{eq:conv_rep}) and (\ref{eq:corr_rep}).}
 	\label{fig:representation}
\end{figure}

Schematically this process is shown in Figure \ref{fig:representation}. The wavefield from the source $\xs \notin D \cup \partial D$ propagates to points $\xr \in \partial D$, from which it is reconstructed at $\x \in D$ using equations (\ref{eq:conv_rep}) and (\ref{eq:corr_rep}). 

Assume that $\wb$ is propagated to $\partial D$ without error, so that $\wb(\xr)$ is exact. If the Green's function $\Gb_D$ is also known exactly, i.e. it is equal to the true Green's function $\Gb^{true}$ then the convolution and correlation representation formulas (\ref{eq:conv_rep}) and (\ref{eq:corr_rep}) reconstruct the same wavefield, which is also the true wavefield $\wb$. If $\Gb_D$ is not known exactly, i.e. $\Gb_D \neq \Gb^{true}$, then both the convolution and correlation wavefields differ from the true field $\wb$ and from each other. Writing this as equations, we have that
\begin{eqnarray}
	\wb^{conv}(\x,\xs,\omega) = \wb^{corr}(\x,\xs,\omega), \;\;\; \mbox{if } \;\; \Gb_D = \Gb^{true} \label{equal}\\
	\wb^{conv}(\x,\xs,\omega) \neq \wb^{corr}(\x,\xs,\omega), \;\;\; \mbox{if } \;\; \Gb_D \neq \Gb^{true} \label{notequal}
\end{eqnarray}
except for pathological cases of symmetry that that are extremely unlikely to occur in real life situations. This is also demonstrated numerically by \cite{Vasconcelos:2016}. In practice, if the fields are redatumed to $\xr$ using e.g. Marchenko redatuming, they may contain errors such as missing and non-physical events \citep{Neut:2015}, as well as errors due to incorrect kinematics of the macro velocity model used for redatuming. In this case, equations (\ref{equal}) and (\ref{notequal}) may hold only approximately.

\subsection{IFWI problem formulation}

The above set-up leads to an FWI problem formulation with an objective function being the $L_2$ norm of the difference between $\wb^{corr}$ and $\wb^{conv}$, where the convolution and correlation fields satisfy constraints (\ref{eq:conv_rep}), (\ref{eq:corr_rep}). 
\subsubsection{PDE constraints}

We set up the inverse problem in the constant density acoustic formulation for pressure $p$ and scaled particle displacement $\ub = \rho \int_t \vb$ in the time domain, so that $\wb(\x,\xs,t) = [p(\x,\xs,t), \ub(\x,\xs,t)]$. Integral equations (\ref{eq:conv_rep}), (\ref{eq:corr_rep}) can be expressed as partial differential equations for pressure and scaled displacement using the forward and adjoint vector-acoustic differential operators \citep{Zheglova:2020} and convolution and correlation areal sources along the boundary $\partial D$:
\begin{eqnarray}
	\Lb(\x,m) \wb^{conv} (\x,\xs,t) &=& \s^{conv} (\xr,t) \delta_{\partial D}(\x,\xr) \label{conv_constr}\\
	\Lbdag (\x,m)  \wb^{corr} (\x,\xs,t) &=& \s^{corr} (\xr,t) \delta_{\partial D}(\x,\xr), \label{corr_constr}
\end{eqnarray}
where:
\begin{description}
	\item $\wb^{i} = [p^{i}, \ub^{i}]$, $i = \left\{conv, corr \right\}$ are the convolution and correlation pressure and particle displacement vector-acoustic wavefields;
	\item $\Lb$ and $\Lbdag$ are the vector-acoustic differential operator and its adjoint:
	\begin{eqnarray*} 
		\Lb(\x,m) = \begin{pmatrix}
			m & \nabla^T \\
			\nabla & \partial_{tt} \mathbf{I}
		\end{pmatrix}, \;\;\;
		\Lbdag(\x,m) = \begin{pmatrix}
			m & -\nabla^T \\
			-\nabla & \partial_{tt} \mathbf{I}
		\end{pmatrix};
	\end{eqnarray*}
	\item $m = 1 / c^2$ is the squared slowness;
	\item $\s^{conv}$ and $\s^{corr}$ are the convolution and correlation sources given at each $\xr \in \partial D$ as
    		\begin{eqnarray*}
    			\s^{conv} (\xr,t) &=& - \Nr(\xr) \wb(\x,\xs,t) = 
    				- \begin{pmatrix}
    				\nb(\xr)^T \ub(\xr,\xs,t) \\
    				\nb(\xr) \; p(\xr,\xs,t)
    			\end{pmatrix}  \\
    			\s^{corr} (\xr,t) &=& - \Kb \Nr(\xr) \wb(\x,\xs,t) = 
    				\begin{pmatrix}
    				- \nb(\xr)^T \ub(\xr,\xs,t) \\
    				\nb(\xr) \; p(\xr,\xs,t)
    			\end{pmatrix} ;
    		\end{eqnarray*}
	\item $\delta_{\partial D}(\x,\xr)$ denotes the areal source distributed along the boundary $\partial D$.
\end{description}
As per the time-domain adjoint-state approach, Equation (\ref{conv_constr}) is solved forward in time, while equation (\ref{corr_constr}) is solved backward in time.

\subsubsection{Objective function}

Based on the reciprocity relations we discuss above, the objective function we use is 
\begin{eqnarray}
	I = \frac{1}{2} \sum_s  \int_0^T dt \int_{D} dV \left\| \Wr \wb^{corr} (\x,\xs,t) - \Wr \wb^{conv} (\x,\xs,t) \right\|_2^2 
	\label{objective}
\end{eqnarray}
where $||\textbf{x}||_2^2=\sum_i x_i^2$ is the squared Euclidean $l_2$ norm and $\Wr$ is the data weighting operator.  In this study we use
\begin{equation}
	\Wr = \begin{pmatrix}
		1 & \zero^T \\
		\zero &  \mathbf{0}
	\end{pmatrix} \label{lambda}
\end{equation}
which samples only pressure. Note that here, unlike the representations in Eqs.~\ref{eq:conv_rep} and~\ref{eq:corr_rep}, we use an explicit time-domain representation for the sake of consistency with our time-domain implementation of the adjoint-state vector-acoustic wave equations.

Equation (\ref{objective}) together with equations (\ref{conv_constr}) and (\ref{corr_constr}) constitute the PDE constrained optimization problem that we solve for the model parameter $m$.

We remark that the convolution and correlation differential equations (\ref{conv_constr}) and (\ref{corr_constr}) together constitute the forward problem that is exactly satisfied by the convolution and correlation fields. The (redatumed) data $\wb(\xr,\xs,t)$ at the boundary $\partial D$ constitute the observed data. The convolution and correlation sources are fully determined by the redatumed data and the geometry of $\partial D$. Unlike in conventional surface FWI and local FWI, we do not directly compare the observed and modelled fields along a line of receivers. Rather, the modelled data are the  pressures ${p^{corr}} = \Wr \wb^{corr}$ and $p^{conv} = \Wr \wb^{conv} $, whose difference must be zero everywhere in the local domain. This process can also be expressed as minimization of $\left\| \begin{pmatrix} \Wr & -\Wr \end{pmatrix} \begin{pmatrix} \wb^{corr} & \wb^{conv} \end{pmatrix}^T \right\|^2_2$. The implications of such an objective function are far wider than it may initially sound: by imposing equivalence between two wavefields, neither of them must be known directly at the location where the objective function in evaluated, allowing us to consider a grid within the entire domain of interest. When only one representation theorem is used \citep[e.g.][]{Cui:2020} the objective function can also be theoretically evaluated at any point in the grid; however, in practical application this would require being able to access the wavefield at such location by means of e.g., Marchenko redatuming, leading to a radical increase in the computational cost associated with the data preparation prior to inversion.


\subsubsection{IFWI gradient}

The gradient of the interferometric objective function with respect to the model parameter $m$ is derived by the adjoint state method \citep{Plessix:2006, Fichtner:2011}. We derive it in Appendix \ref{AppA}, and show here only the result. In the general form, the gradient of the interferometric objective function with respect to the squared slowness $m$ is given by
\begin{eqnarray}
	\frac{\partial I}{\partial m} &=& \int_0^T(-\wb^{corr,T} \Wr \wb^{corr \dagger} - \wb^{conv, T}  \Wr \wb^{conv \dagger}) dt \nonumber
\end{eqnarray}
where $\wb^{conv \dagger} = \begin{pmatrix}p^{conv \dagger}\\ \ub^{conv \dagger}\end{pmatrix}$ and $\wb^{corr \dagger} = \begin{pmatrix} p^{corr \dagger}\\ \ub^{corr \dagger} \end{pmatrix}$ are the adjoint convolution and correlations wavefields that satisfy the adjoint  convolution and correlation equations:
\begin{eqnarray}
    \Lbdag \wb^{conv \dagger} &=& - \Wradj \Wr (\wb^{corr} - \wb^{conv}) \label{adj_conv} \\
    \Lb \wb^{corr \dagger} &=& \Wradj \Wr (\wb^{corr} - \wb^{conv}) \label{adj_corr},
\end{eqnarray}
Equations (\ref{adj_conv}) and (\ref{adj_corr}) are solved respectively backward and forward in time.

With the choice of $\Wr$ as in (\ref{lambda}), the gradient becomes
\begin{eqnarray}	
	\frac{\partial I}{\partial m} & = & \int_0^T (-p^{corr} p^{corr \dagger} - p^{conv} p^{conv \dagger}) dt, \label{gradient}
\end{eqnarray}

Since the residual $\wb^{corr} - \wb^{conv}$ is calculated over the whole $D$, the convolution and correlation adjoint sources, i.e. right-hand sides of equations (\ref{adj_conv}) and (\ref{adj_corr}) are volume sources injected everywhere in the local domain. This is different from conventional FWI, where the adjoint sources are usually injected at the receiver locations. Similar to conventional FWI, the adjoint fields are computed in the opposite time direction to the forward problem, and zero-lag cross-correlations of the forward and adjoint fields are computed to obtain the gradient.

We implement the iterative inversion by combining the IFWI objective and gradient with the L-BFGS optimization method.

\subsubsection{Convexity of the interferometric objective function}

In this section, we investigate the convexity of the interferometric objective function in the target domain using exact data, and compare it to the surface FWI objective function. We consider an example, where we obtain the true model, shown in Figure 2, by applying a mask to a modified Marmousi model. We place six sources between 1 and 3 km in the horizontal direction at a depth of 0.1 km and use a Ricker wavelet with a peak frequency of 15 Hz as the source signature.

\begin{figure}\centering
 	\subfigure[]{\includegraphics[width = 0.74\textwidth,trim={0cm 2cm 0cm 2cm},clip]{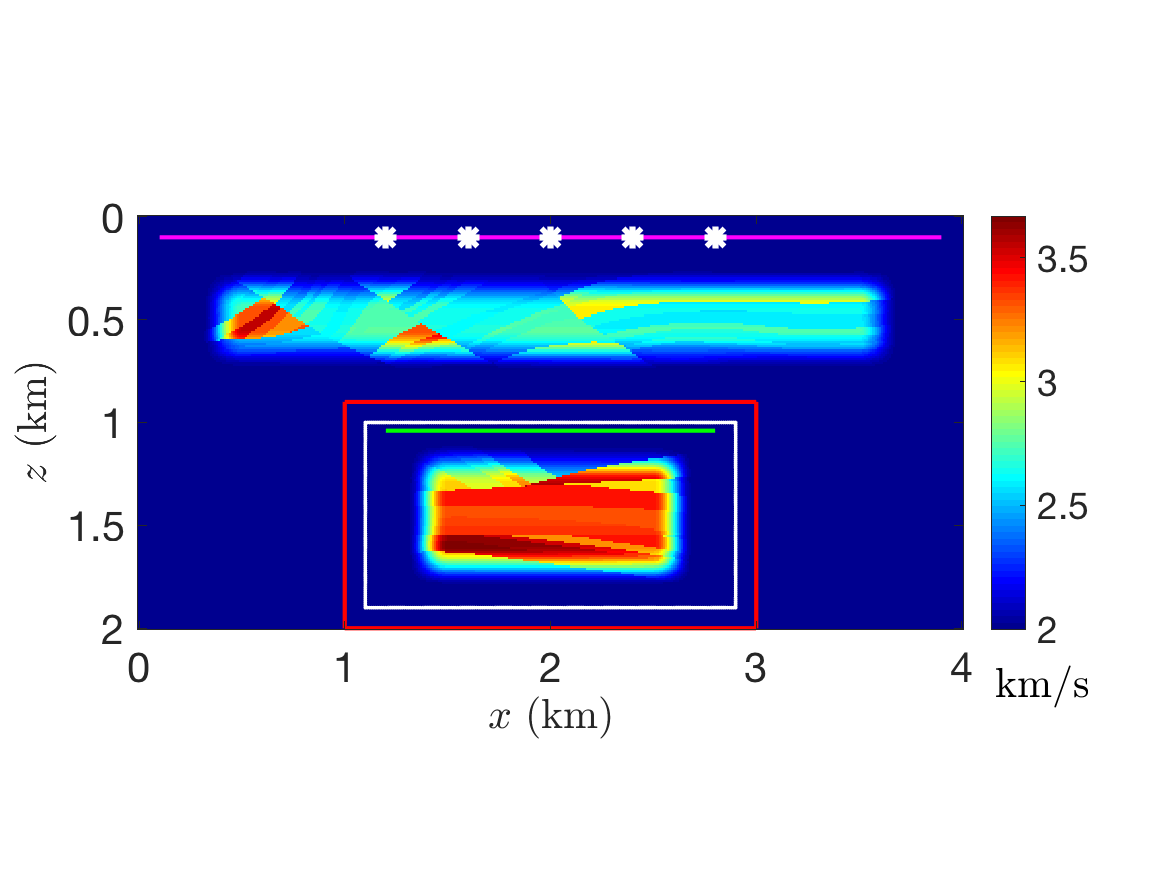}} 
	\\
  	\subfigure[]{\includegraphics[width = 0.4\textwidth,trim={0cm 1.4cm 0cm 3cm},clip]{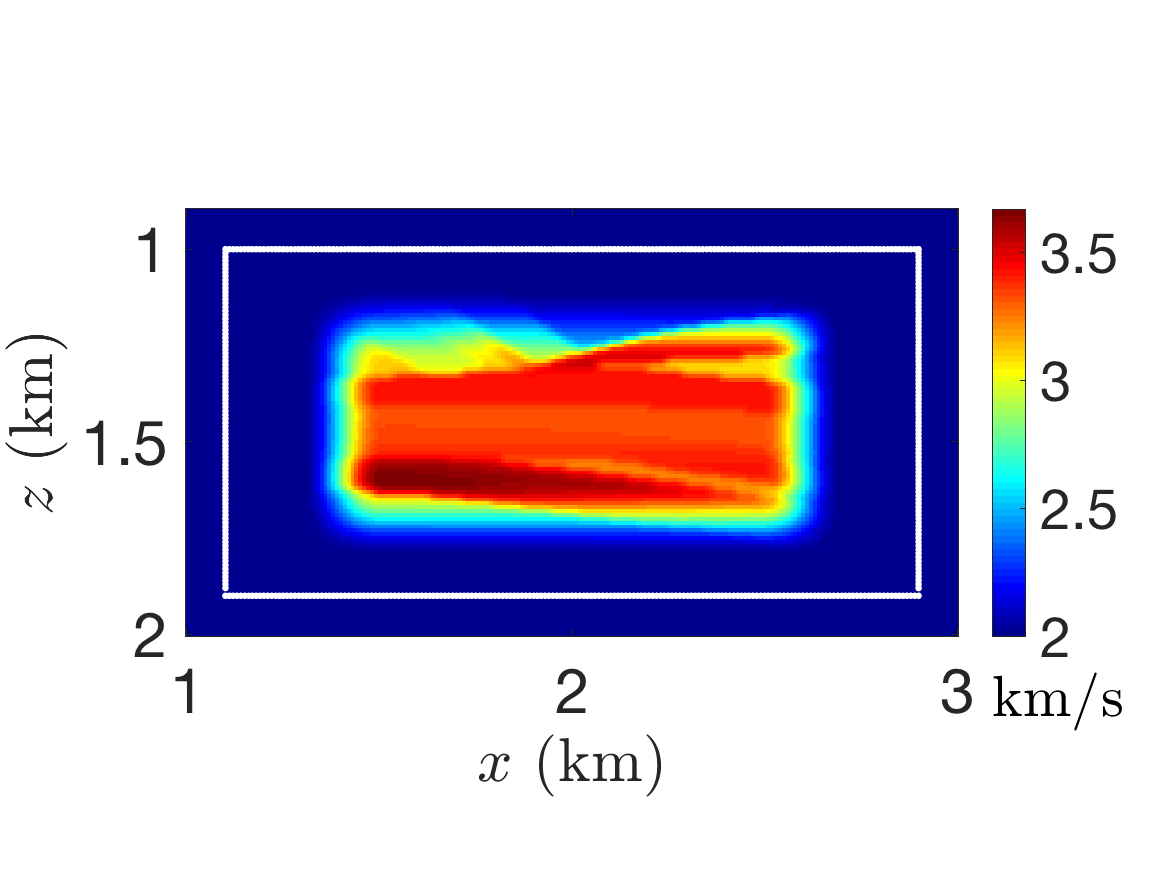}}
 	\subfigure[]{\includegraphics[width = 0.4\textwidth,trim={0cm 1.4cm 0cm 3cm},clip]{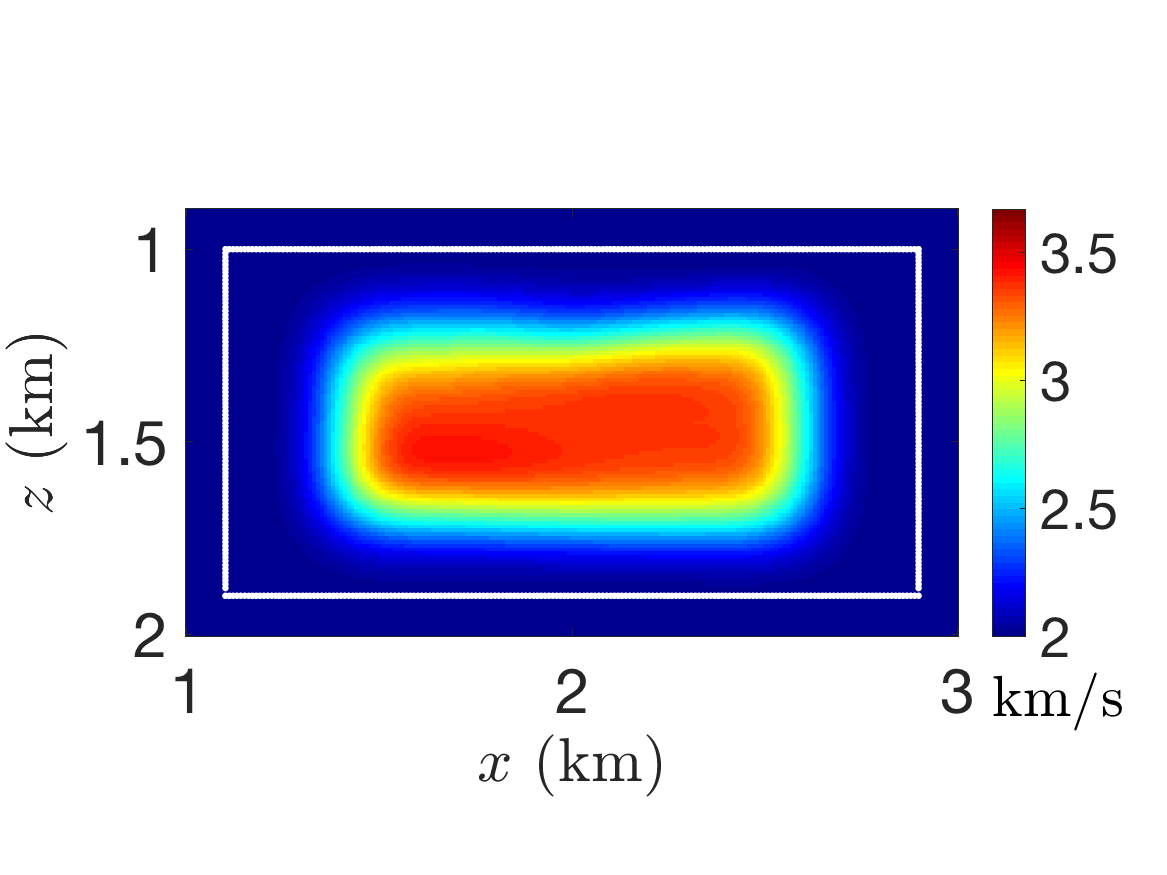}}
	\caption{(a) True model for objective function behaviour demonstration. White stars represent surface sources. The red line shows the local subdomain, the white line shows the injection boundary $\partial D$. The magenta line shows the surface receivers. (b), (c) Two starting velocity models used for objective and gradient behaviour analysis, obtained by applying a Gaussian filter with STD of (a) 10 m and (b) 80 m to the local domain of the true model.}
 	\label{fig:true_model}
\end{figure}

\begin{figure}
	\centering
	\includegraphics[width = 0.5\textwidth,trim={0cm 0cm 0cm 0cm},clip]{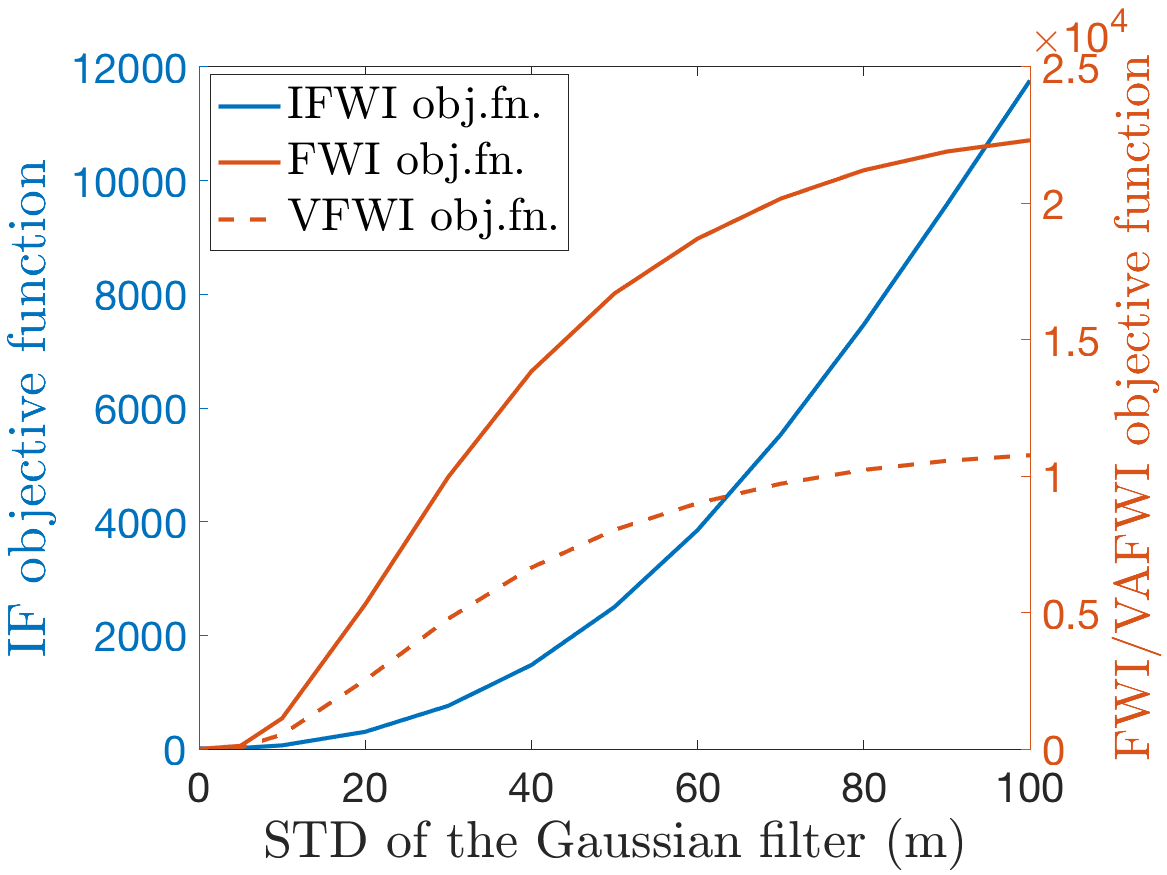}
 	\caption{Objective function behaviour as function of smoothness of the target velocity model. The IFWI objective function is displayed in blue, whilst the classical and vector-acoustic FWI objective functions are shown in solid and dashed red, respectively.} 
 	\label{fig:obj_smooth_trend}
\end{figure}

Because a kinematically accurate macro velocity model is usually needed for both full waveform inversion and redatuming, we are primarily interested in the response of the objective function to the lack of high wavenumber components in the model. Therefore, we apply 2D Gaussian smoothing to the target area, where we vary the standard deviation of the Gaussian filter from 5 to 100 m to obtain progressively smoother models of the target, and compute the objective function value for each of the smoothed models. Examples of such smoothed local models are shown in Figure \ref{fig:true_model} (b), (c). Figure \ref{fig:obj_smooth_trend} shows the interferometric objective function variation with the STD of the Gausian filter. For comparison, we show the surface FWI objective and vector-acoustic FWI (VAFWI) objective function \citep{Zheglova:2020} for the same smoothed models. To compute the FWI and VAFWI objective, receivers are placed every 10 m between 0.2 and 3.8 km at depth 0.1 km. For the calculation of FWI and VAFWI objective function, we apply no smoothing to the overburden layer. Note that, apart from the difference in absolute values, the interferometric objective function is more convex than both the FWI and VAFWI objective functions, in addition, it has a wider basin of attraction near the global minimum. Based on this experiment, we can expect the IFWI to exhibit fast initial convergence with a subsequent slow-down near the minimum. 
We can also expect the IFWI to be more resistant to errors in the initial model, which we show to be case with \textit{exact} data.

\subsubsection{Detailed analysis of the interferometric gradient}

In this section, we take a closer look at the gradient of the interferometric objective function. 

First, we observe that each of the two terms in the gradient (equation \ref{gradient}) is a zero-lag correlation of the forward field with the corresponding adjoint field. Each of the adjoint fields in turn consists of two parts: a part due to the convolution forward field, and a part due to the correlation forward field, as follows from the adjoint equations (\ref{adj_conv}) and (\ref{adj_corr}). We denote these adjoint field components as follows:
\begin{enumerate}
  	\item \label{convconv} $\wb^{conv \dagger}_{conv} = \Gamma^\dagger \Wr^{\dagger} \Wr \Gamma w^{conv}$ is the part of the convolution adjoint field $\wb^{conv \dagger}$ due to the convolution part of the adjoint source $\Wradj \Wr \wb^{conv}$  	
	\item \label{convcorr} $\wb^{conv \dagger}_{corr} = - \Gamma^\dagger \Wr^{\dagger} \Wr \Gamma^\dagger w^{corr}$ is the part of the convolution adjoint field $\wb^{conv \dagger}$ due to the correlation part of the adjoint source $- \Wradj \Wr \wb^{corr}$ 
	\item \label{corrcorr} $\wb^{corr \dagger}_{corr} = \Gamma \Wr^{\dagger} \Wr \Gamma^\dagger w^{corr}$ is the part of the correlation adjoint field $\wb^{corr \dagger}$ due to the correlation part of the adjoint source $\Wradj \Wr \wb^{corr}$, and 
	\item \label{corrconv} $\wb^{corr \dagger}_{conv} = \Gamma \Wr^{\dagger} \Wr \Gamma w^{corr}$ is the part of the correlation adjoint field $\wb^{corr \dagger}$ due to the convolution part of the adjoint source $- \Wradj \Wr \wb^{conv}$, 
\end{enumerate} 
where $\Gamma$ and $\Gamma^{\dagger}$ are Green's functions for operators $\Lb$ and $\Lb^{\dagger}$. 
Correspondingly, the four terms in the gradient are: 
\begin{eqnarray}
	\frac{\partial I}{\partial m}^{(i)} & = & - \int_0^T p^{conv} p^{conv \dagger}_{conv} dt\\
	\frac{\partial I}{\partial m}^{(ii)} & = & - \int_0^T p^{conv} p^{conv \dagger}_{corr} dt\\
	\frac{\partial I}{\partial m}^{(iii)} & = & - \int_0^T p^{corr} p^{corr \dagger}_{corr} dt\\
	\frac{\partial I}{\partial m}^{(iv)} & = & - \int_0^T p^{corr} p^{corr \dagger}_{conv} dt
\end{eqnarray}
  
We use two starting models to analyze the behaviour of the IFWI gradient, shown in Figure \ref{fig:true_model} (b) and (c). The model in Figure \ref{fig:true_model} (b) represents late inversion stages, while the model in Figure \ref{fig:true_model} (c) represents early inversion stages. The original field is generated by a Ricker wavelet source with a peak frequency of 15 Hz located at the middle top of the model (Figure \ref{fig:true_model}) at $z = 0.1$ km, $x = 2$ km. 

First, we look at the gradient obtained in the starting model in Figure \ref{fig:true_model} (b). This model is very close to the true model and contains almost all of the high wavenumber details in the local subdomain. 
  
\begin{figure}\centering
 	\subfigure[]{\includegraphics[width = 0.4\textwidth,trim={0cm 1.5cm 0cm 2cm},clip]{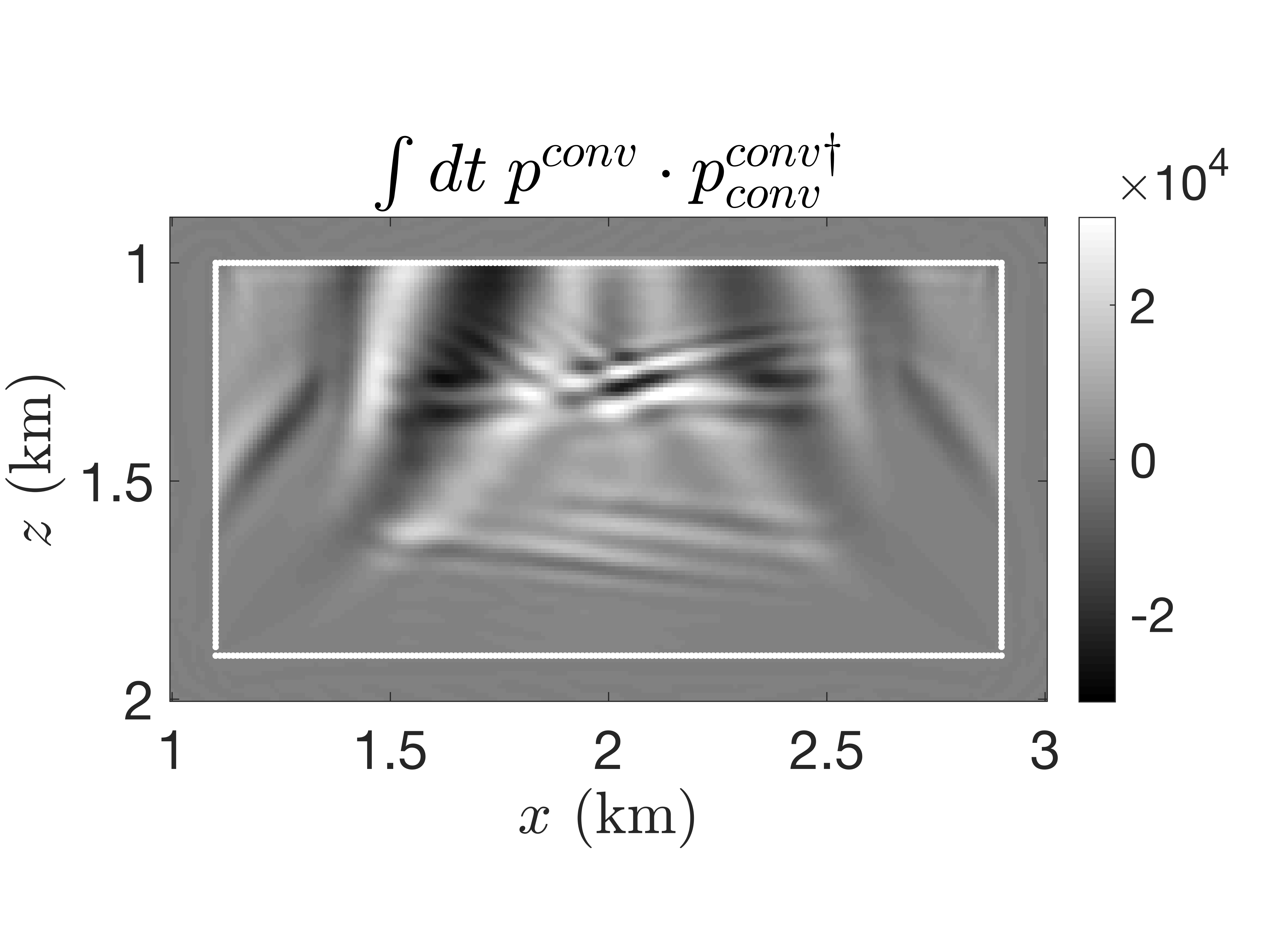}}
  	\subfigure[]{\includegraphics[width = 0.4\textwidth,trim={0cm 1.5cm 0cm 2cm},clip]{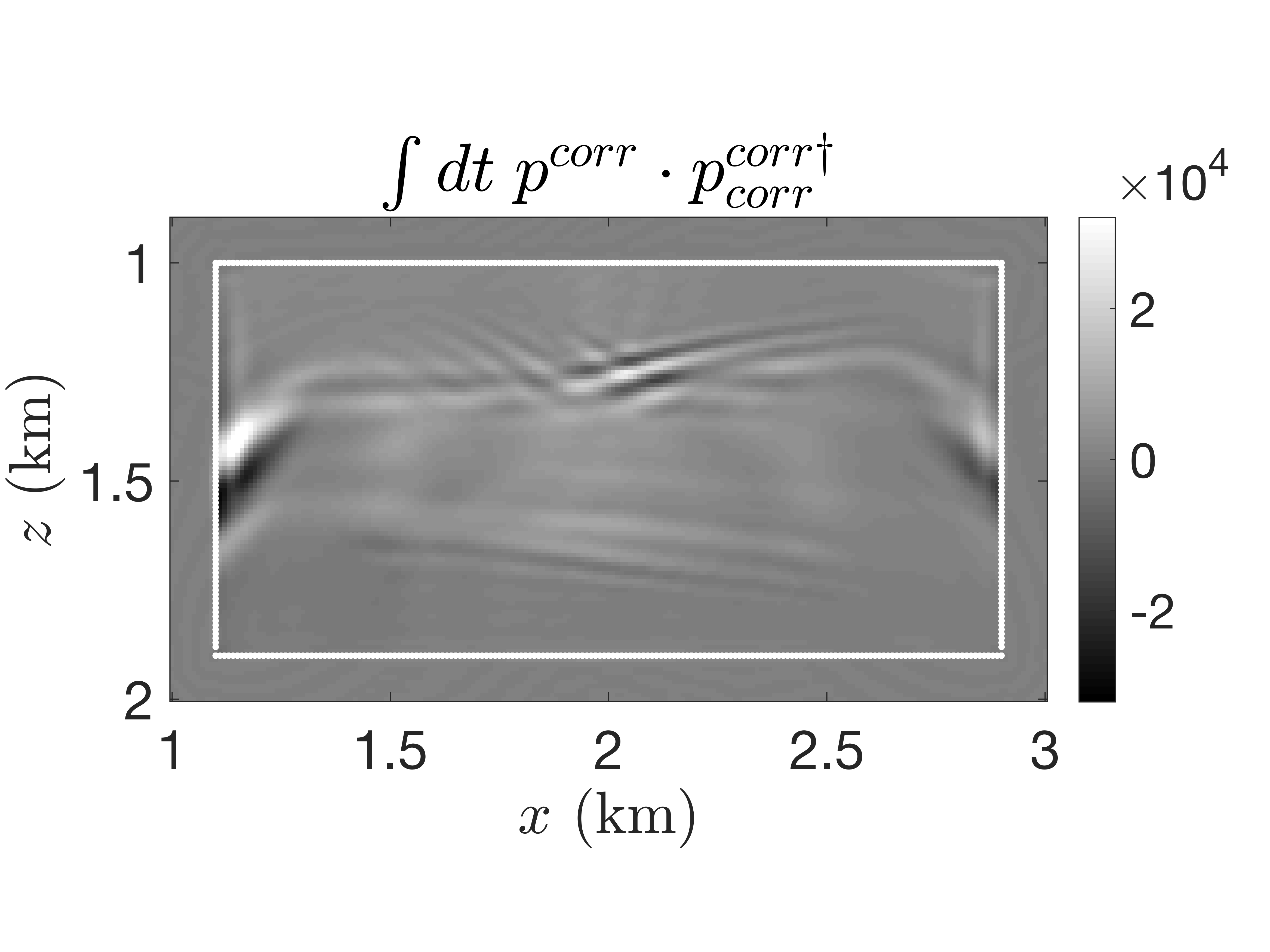}}
	\subfigure[]{\includegraphics[width = 0.4\textwidth,trim={0cm 1.5cm 0cm 2cm},clip]{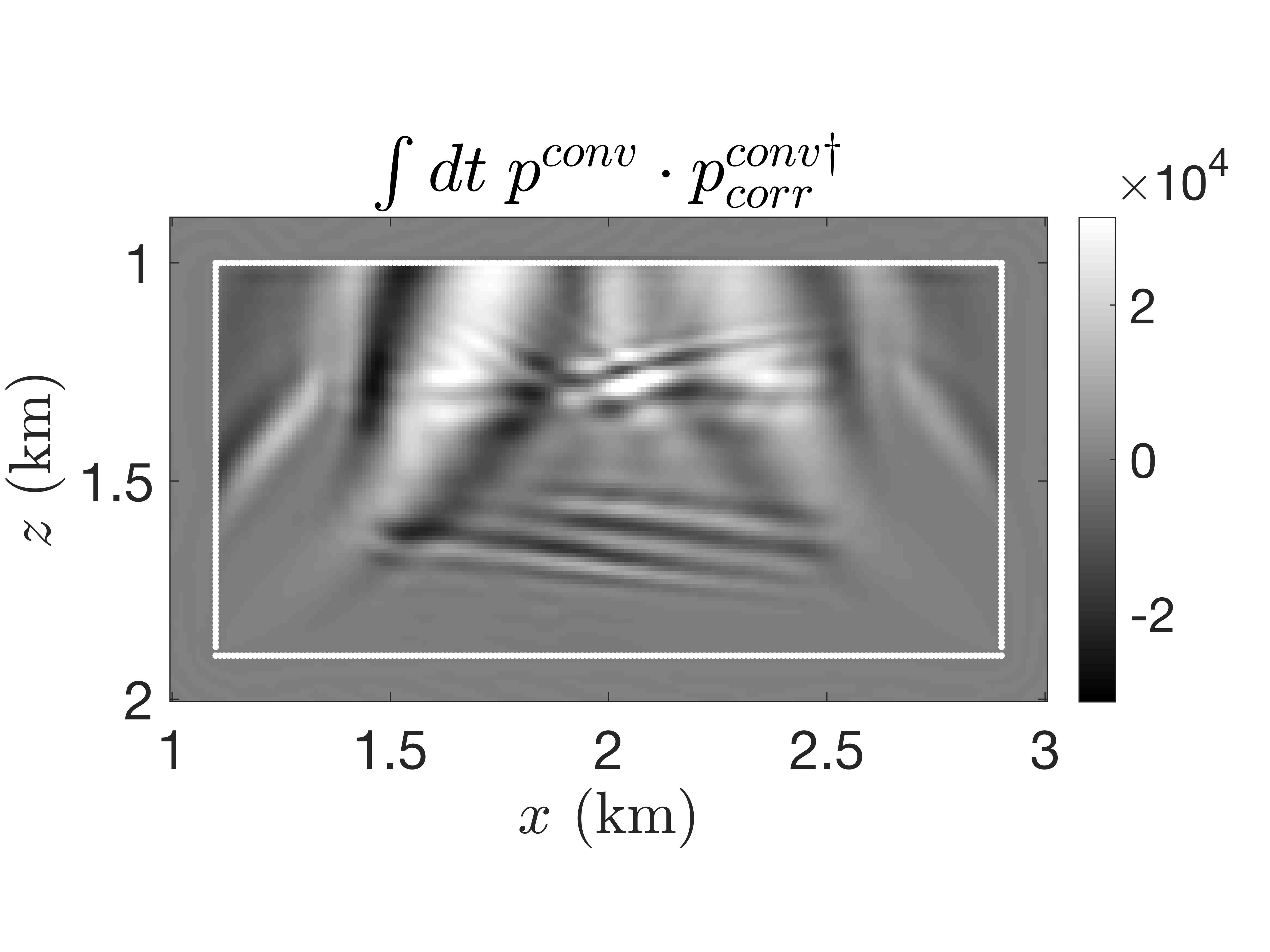}}
  	\subfigure[]{\includegraphics[width = 0.4\textwidth,trim={0cm 1.5cm 0cm 2cm},clip]{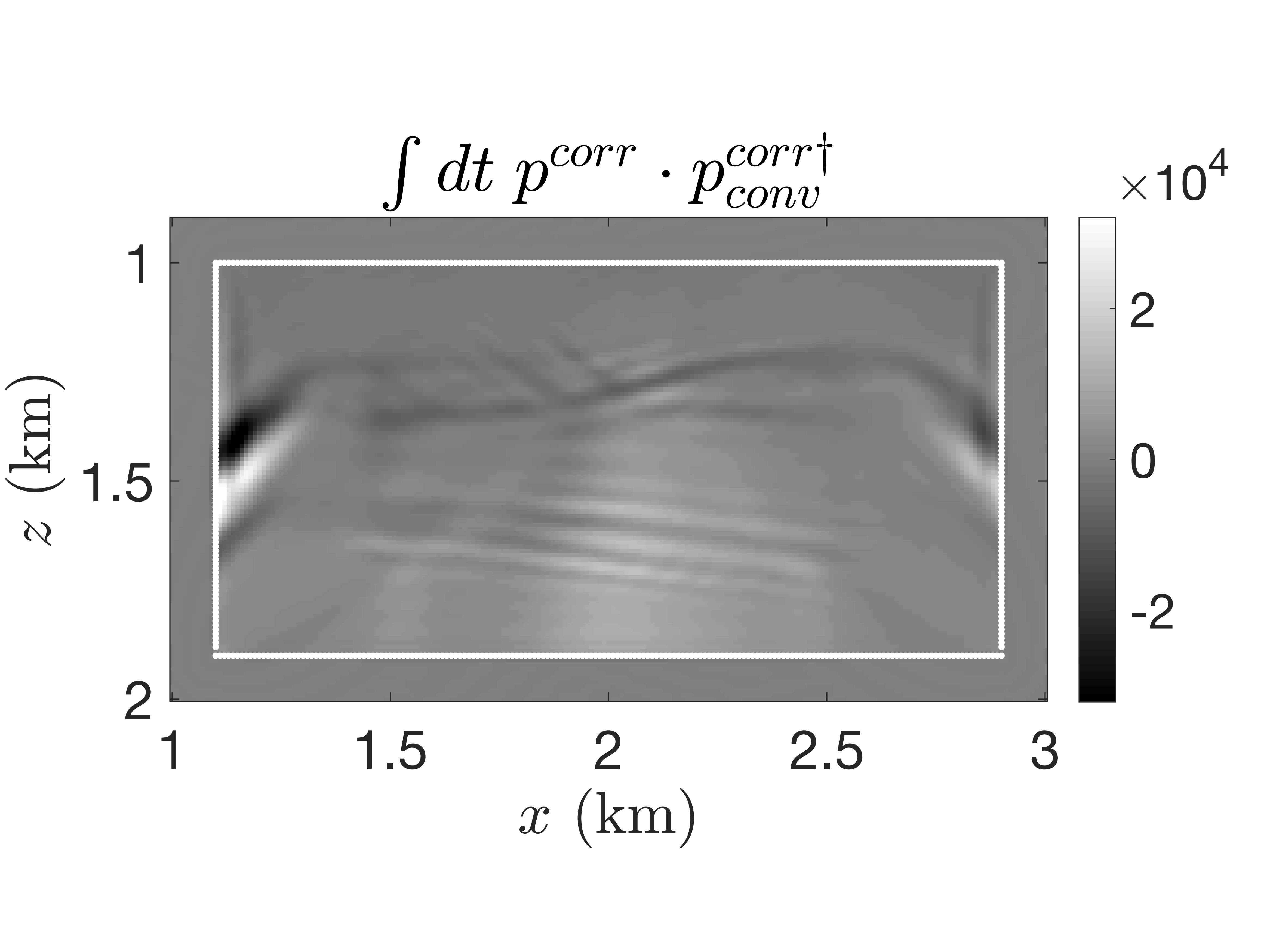}}
	\subfigure[]{\includegraphics[width = 0.4\textwidth,trim={0cm 1.5cm 0cm 2cm},clip]{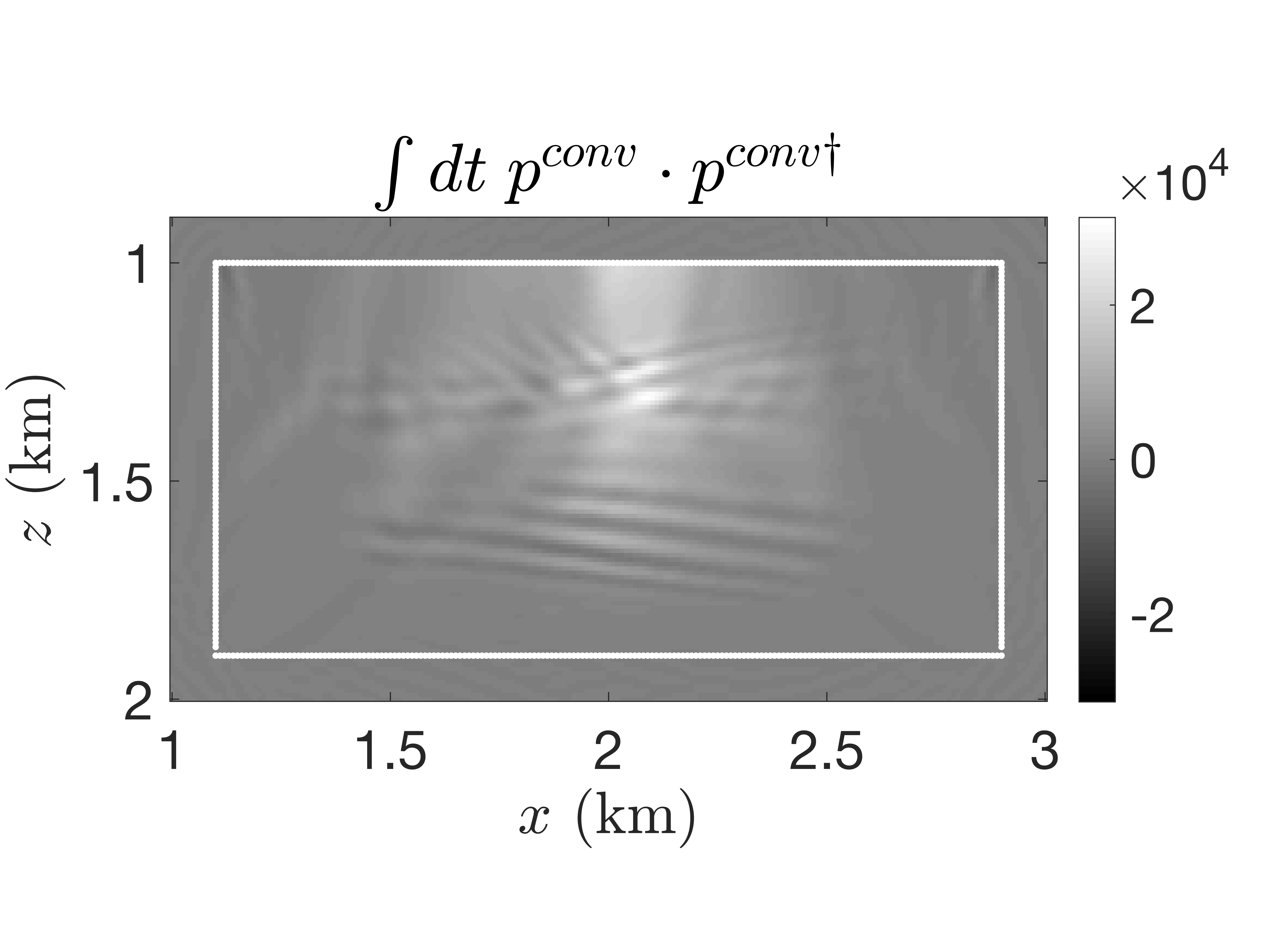}}
 	\subfigure[]{\includegraphics[width = 0.4\textwidth,trim={0cm 1.5cm 0cm 2cm},clip]{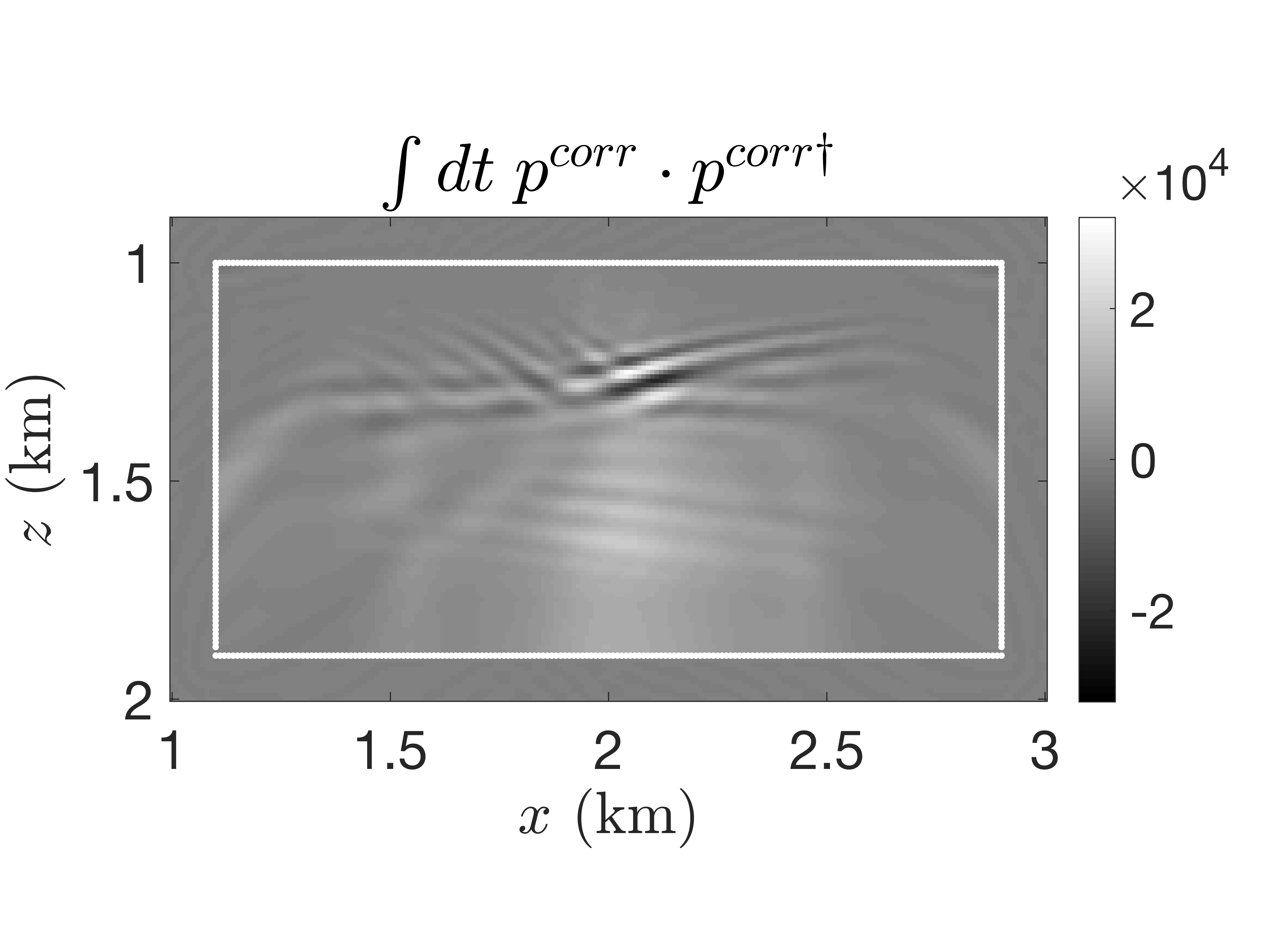}}
 	\caption{Zero-lag correlation components of the gradient for the true and starting models shown in Figure \ref{fig:true_model} (a) and (b), for $t = 0$ to 4 s. (a) $\frac{\partial I}{\partial m}^{(i)}$; (c) $\frac{\partial I}{\partial m}^{(ii)}$; (e) $\frac{\partial I}{\partial m}^{(i)}+\frac{\partial I}{\partial m}^{(ii)}$; (b) $\frac{\partial I}{\partial m}^{(iii)}$; (d) $\frac{\partial I}{\partial m}^{(iv)}$; (f) $\frac{\partial I}{\partial m}^{(iii)}+\frac{\partial I}{\partial m}^{(iv)}$.}
 	\label{fig:conv_components}
\end{figure}

Figures \ref{fig:conv_components} (a) and (c) show the individual cross-correlation terms $\frac{\partial I}{\partial m}^{(i)}$ and $\frac{\partial I}{\partial m}^{(ii)}$ of the gradient, and Figure \ref{fig:conv_components} (e) shows their sum. The two terms $\frac{\partial I}{\partial m}^{(i)}$ and $\frac{\partial I}{\partial m}^{(ii)}$ in Figure \ref{fig:conv_components} (a) and (c) have reversed polarity in both low and high wavenumber components, which happens because the convolution and correlation forward fields in the adjoint sources have opposite signs. Due to the mismatch in the convolution and the correlation forward fields, these cross-correlation terms with opposite polarity are not exactly aligned and do not sum to zero, producing a pattern in the gradient that closely follows the features in the model perturbation. The same can be observed to a certain degree for the individual cross-correlation terms $\frac{\partial I}{\partial m}^{(iii)}$ and $\frac{\partial I}{\partial m}^{(iv)}$ shown in Figure \ref{fig:conv_components} (b) and (d) and their sum shown in Figure \ref{fig:conv_components} (f), although it is apparent that other mechanisms are also present in the correlation update. Thus, Figure \ref{fig:conv_components} visually demonstrates that when the initial model is close to the starting model, the gradient update is driven by the mismatch in the convolution and correlation forward wavefields at the later stages of the inversion. 

\begin{figure}\centering
 	\subfigure[]{\includegraphics[width = 0.4\textwidth,trim={0cm 1.5cm 0cm 2cm},clip]{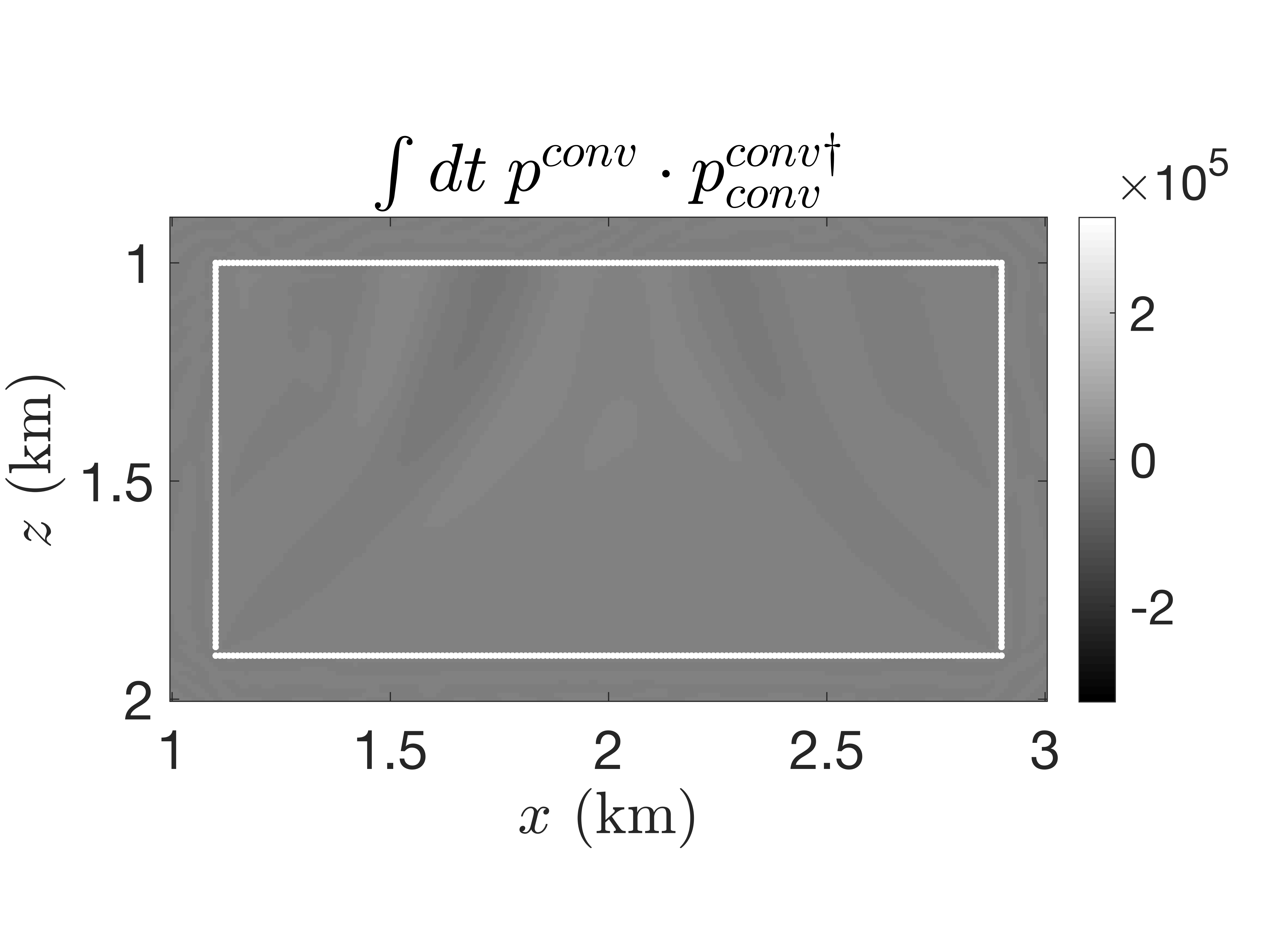}} 
  	\subfigure[]{\includegraphics[width = 0.4\textwidth,trim={0cm 1.5cm 0cm 2cm},clip]{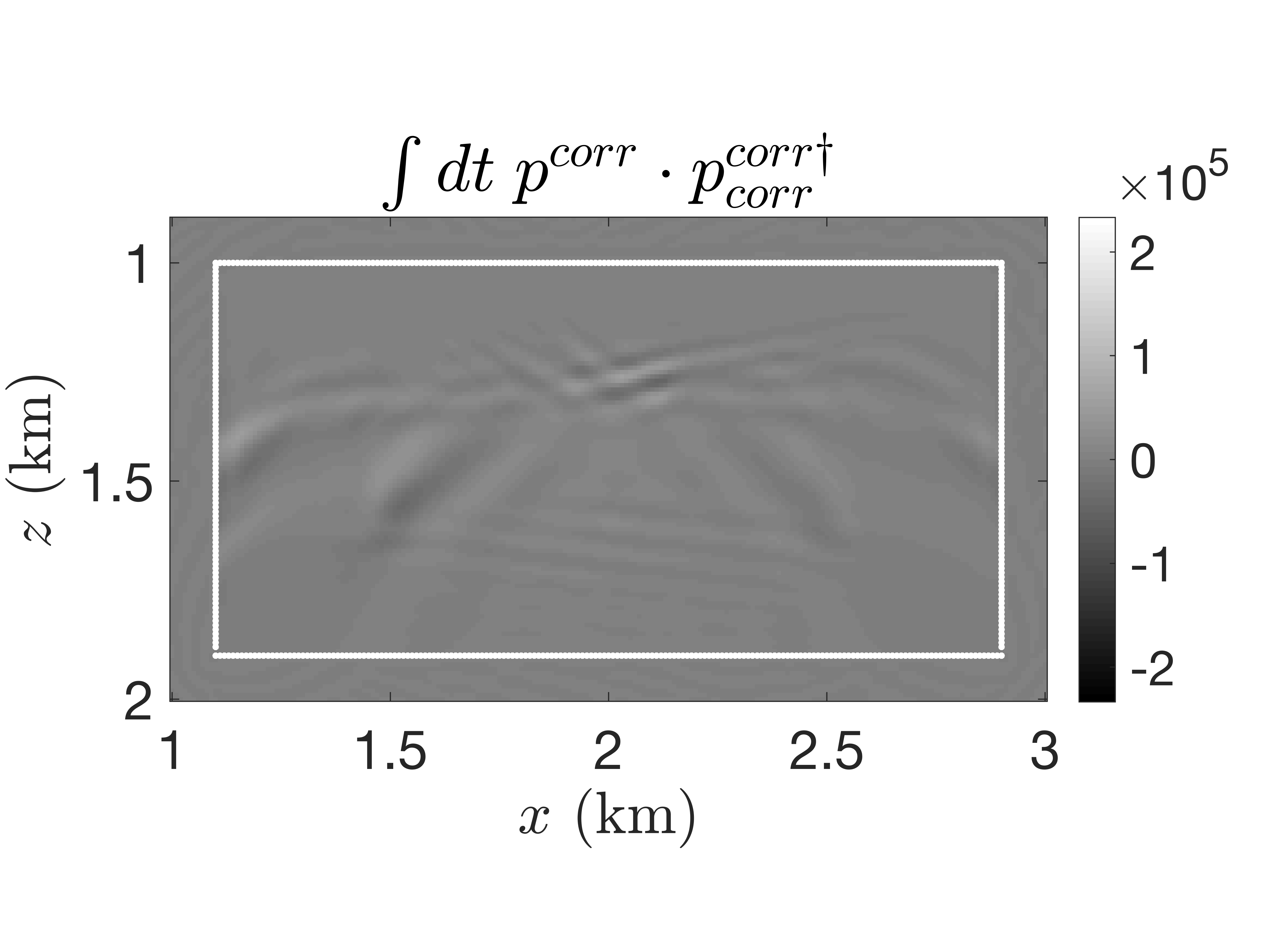}} 
	\subfigure[]{\includegraphics[width = 0.4\textwidth,trim={0cm 1.5cm 0cm 2cm},clip]{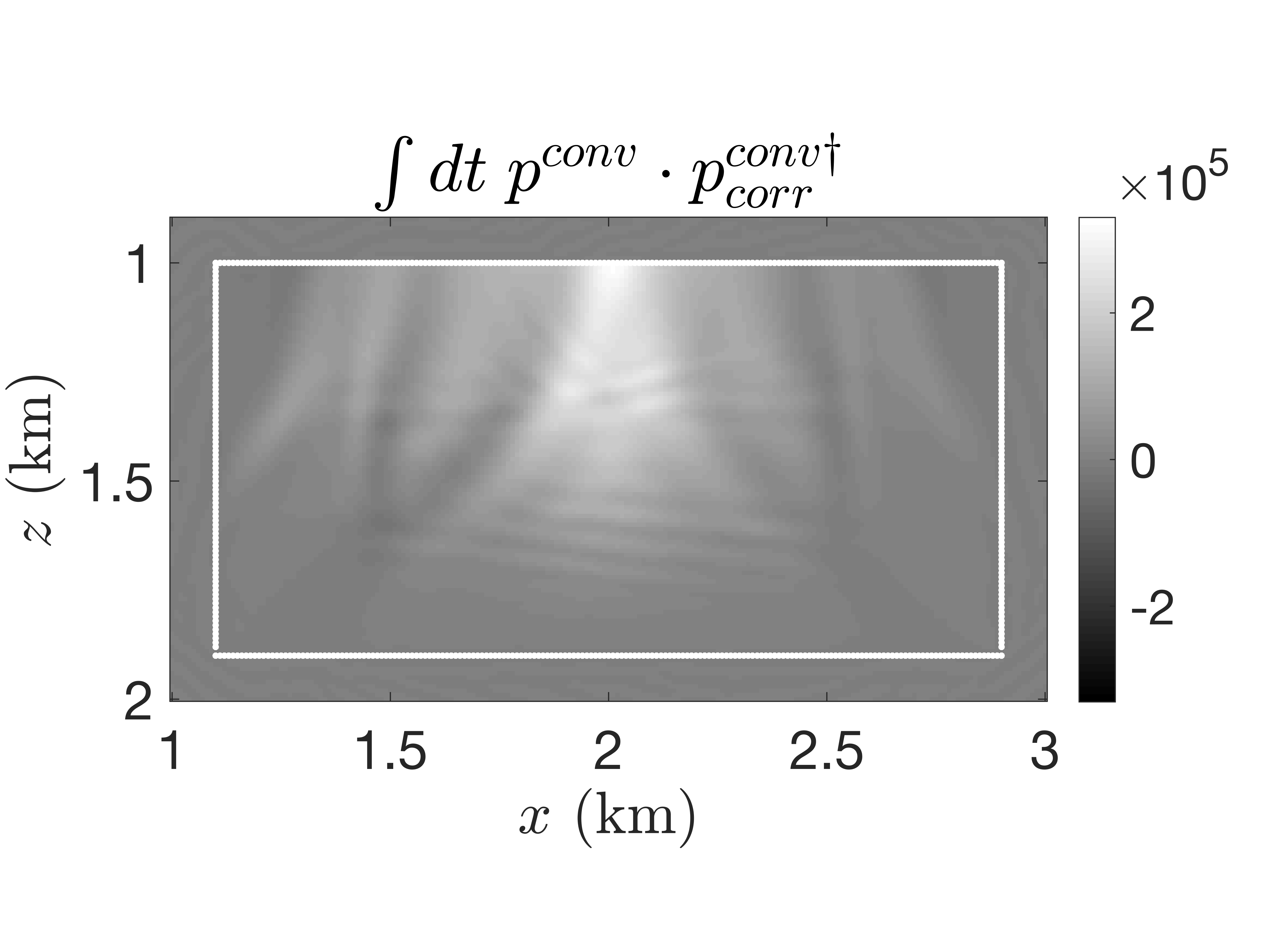}} 
  	\subfigure[]{\includegraphics[width = 0.4\textwidth,trim={0cm 1.5cm 0cm 2cm},clip]{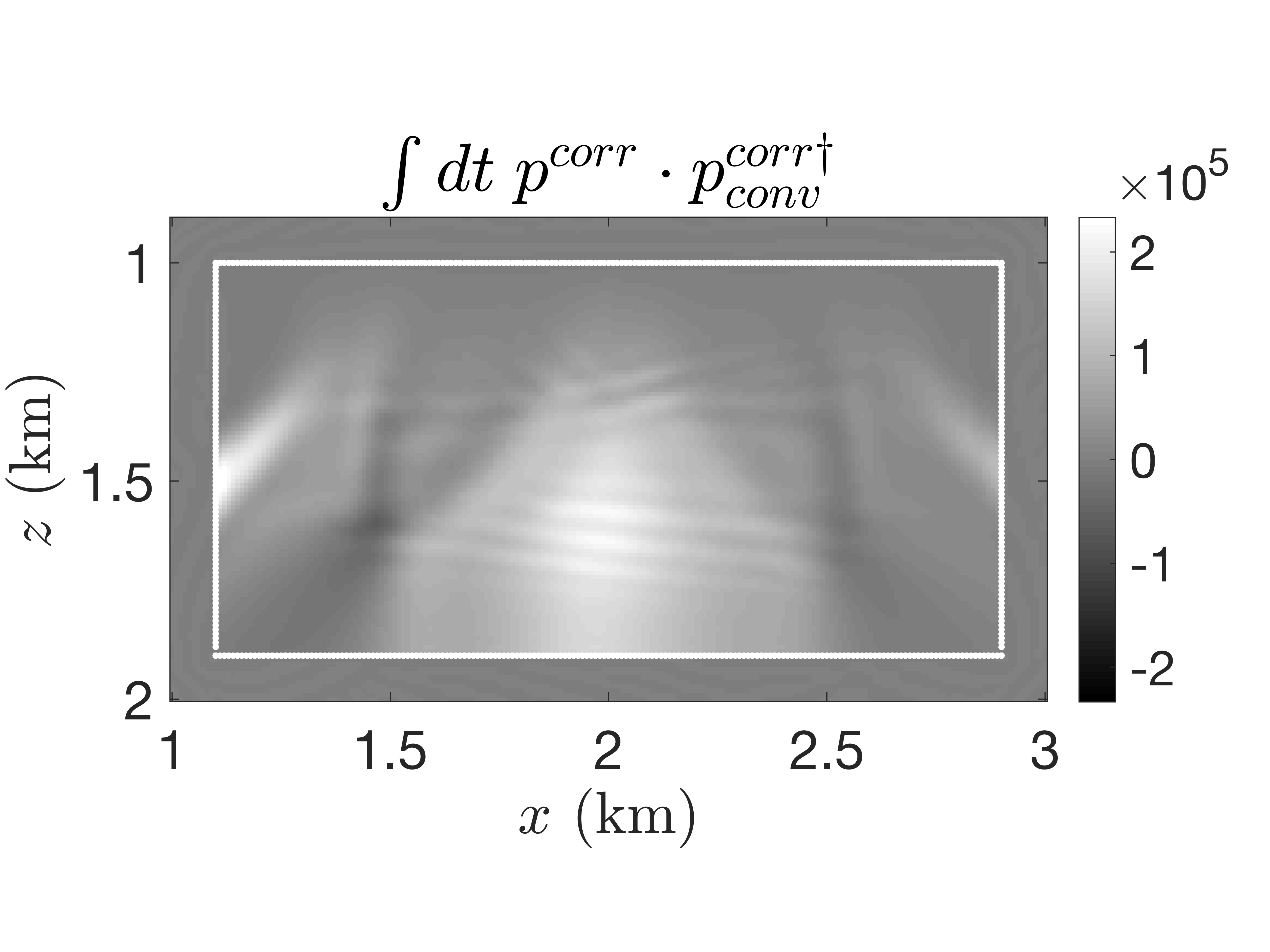}} 
	\subfigure[]{\includegraphics[width = 0.4\textwidth,trim={0cm 1.5cm 0cm 2cm},clip]{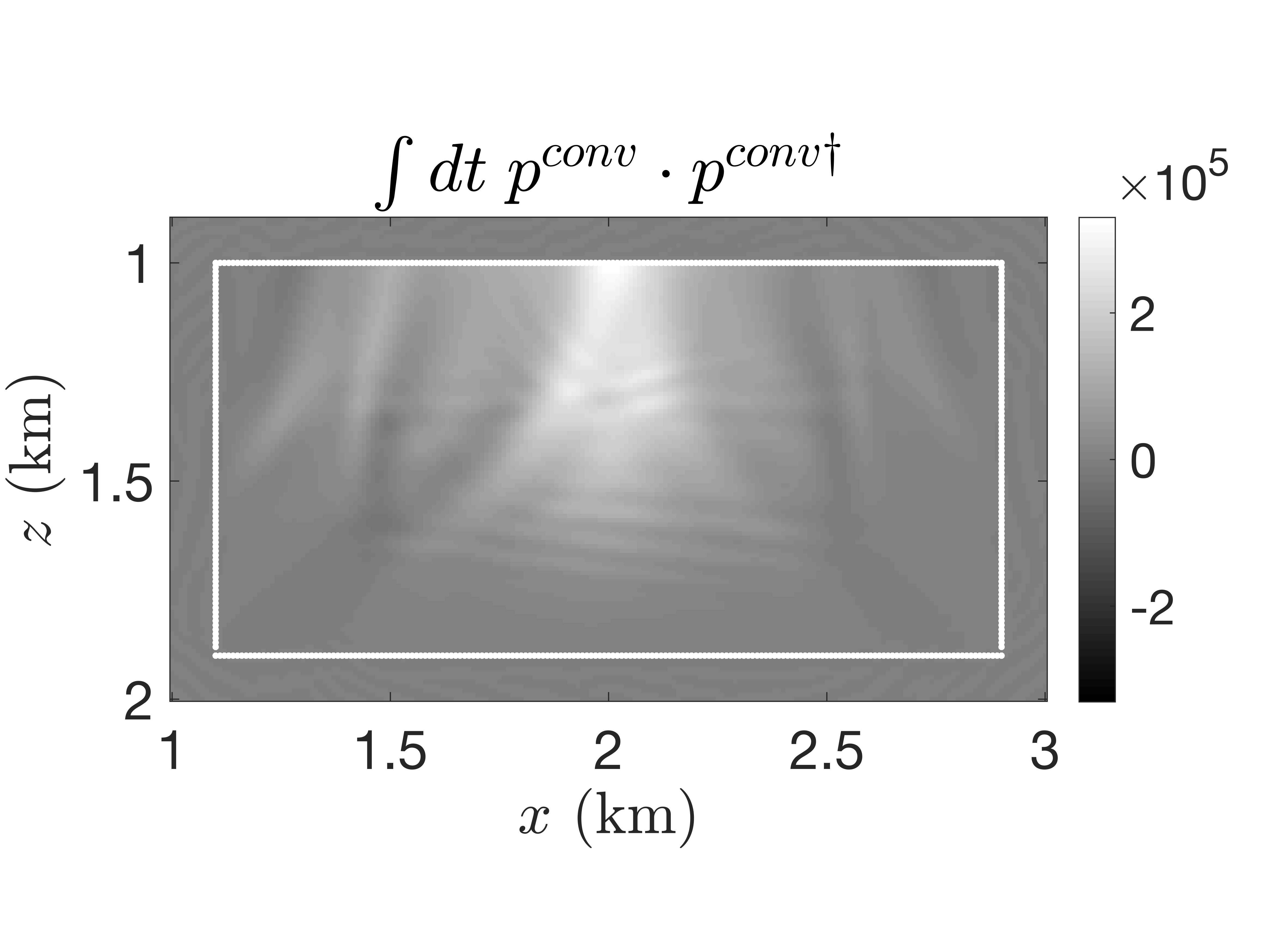}} 
 	\subfigure[]{\includegraphics[width = 0.4\textwidth,trim={0cm 1.5cm 0cm 2cm},clip]{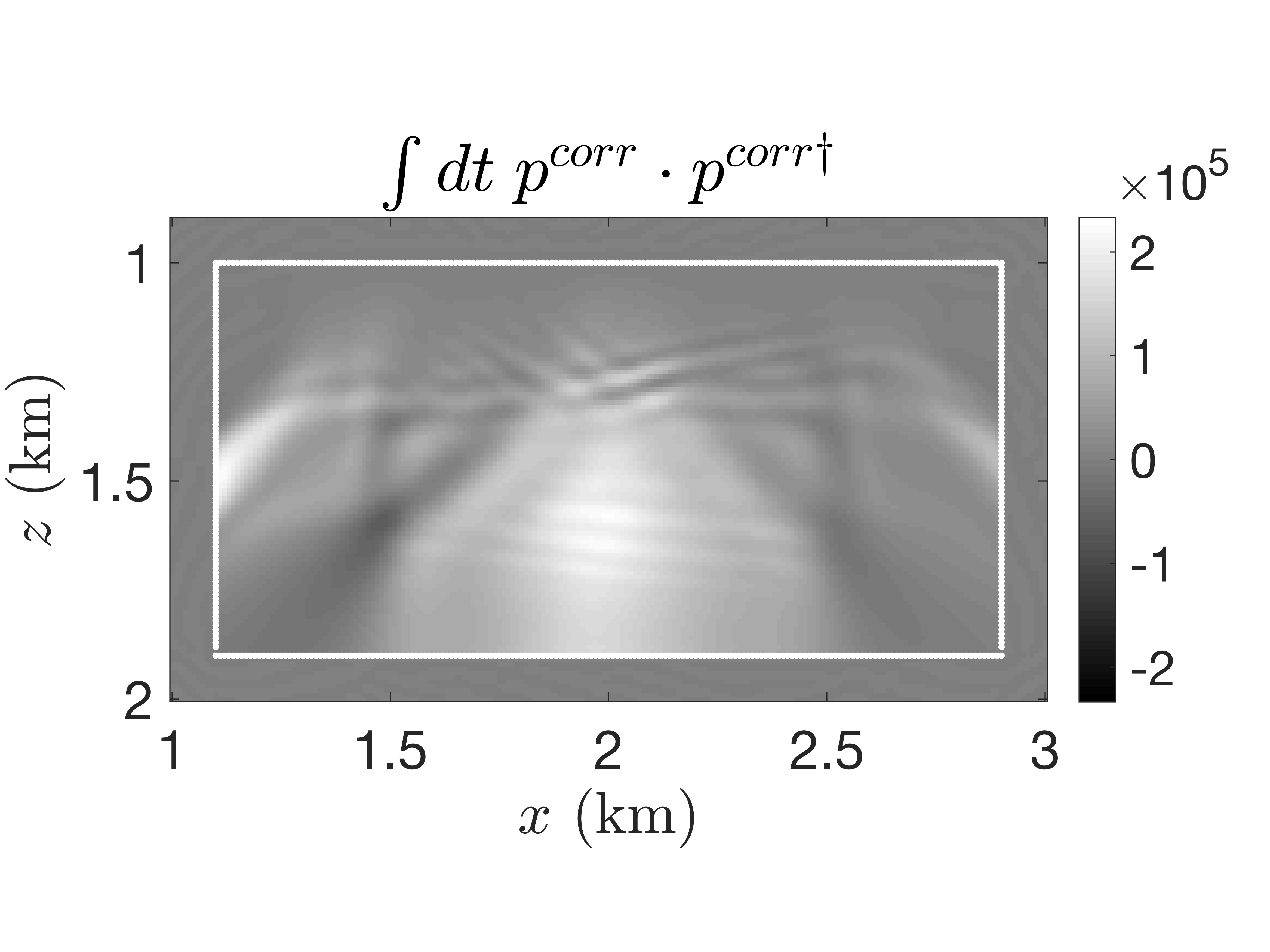}} 
 	\caption{Zero-lag correlation components of the gradient for the true and starting models shown in Figure \ref{fig:true_model} (a) and (c), for $t = 0$ to 4 s.  (a) $\frac{\partial I}{\partial m}^{(i)}$; (c) $\frac{\partial I}{\partial m}^{(ii)}$; (e) $\frac{\partial I}{\partial m}^{(i)}+\frac{\partial I}{\partial m}^{(ii)}$; (b) $\frac{\partial I}{\partial m}^{(iii)}$; (d) $\frac{\partial I}{\partial m}^{(iv)}$; (f) $\frac{\partial I}{\partial m}^{(iii)}+\frac{\partial I}{\partial m}^{(iv)}$.}
 	\label{fig:conv_components8}
\end{figure}

The situation is somewhat different for the starting model shown in Figure \ref{fig:true_model} (c). The individual cross-correlation terms and their sums for this model are shown in Figure \ref{fig:conv_components8}. This model is very smooth and lacks any high-wavenumber details, so the convolution forward field does not undergo any scattering inside the local domain and is purely down-going: $p^{conv} \downarrow$. It also generates a purely down-going adjoint field $p^{conv \dagger}_{conv} \downarrow$ and a purely up-going field $p^{corr \dagger}_{conv} \uparrow$. Zero-lag correlation of the fields traveling in the same direction generates low-wavenumber updates, whereas zero-lag correlation of the fields travelling in opposite directions generates low-wavenumber updates. The term $\frac{\partial I}{\partial m}^{(i)}$ is a cross-correlation of two down going waves, and therefore updates only low wavenumbers, Figure \ref {fig:conv_components8} (a). The correlation field $p^{corr}$ has both transmitted and reflected waves from the true model that are respectively down- and up-going: $p^{corr} \downarrow + \uparrow$, and generates up- and down-going waves in the adjoint field components $p^{conv \dagger}_{corr}: \uparrow + \downarrow$ and $p^{corr \dagger}_{corr}: \uparrow + \downarrow$. Consequently, $\frac{\partial I}{\partial m}^{(ii)}$ contains zero-lag correlations of up- and down-going waves as well as down- and down-going waves: $\uparrow \downarrow + \downarrow \downarrow$ and therefore updates both low and high wavenumbers, Figure \ref {fig:conv_components8} (c). Likewise, the term $\frac{\partial I}{\partial m}^{(iv)}$ contains cross-correlations of the up- and up-going waves as well as up- and down-going waves $\uparrow \uparrow + \uparrow \downarrow$, Figure \ref {fig:conv_components8} (d). Finally $\frac{\partial I}{\partial m}^{(iii)}$ contains all possible cross-correlations: $\uparrow \uparrow + \uparrow \downarrow +  \downarrow \uparrow + \downarrow \downarrow$ and updates all wavenumbers to some degree, Figure  \ref {fig:conv_components8} (b). 

The second observation we make from Figures \ref {fig:conv_components} (e), (f) and \ref {fig:conv_components8} (e), (f) is that the convolution and correlation components in the gradient contribute to different parts of the model update. The convolution term $\int_0^T p^{conv} p^{conv \dagger} dt$, Figure \ref{fig:conv_components} and \ref{fig:conv_components8} (c), updates the target area mostly above the bottom reflector of the target. The correlation term $\int_0^T p^{corr} p^{corr \dagger} dt$, Figure  \ref{fig:conv_components} and \ref{fig:conv_components8} (f), updates the target area mostly from the top reflector to the bottom of the injection boundary $\partial D$. 

\begin{figure}\centering
 	\subfigure[]{\includegraphics[width = 0.4\textwidth,trim={0cm 1.5cm 0cm 3cm},clip]{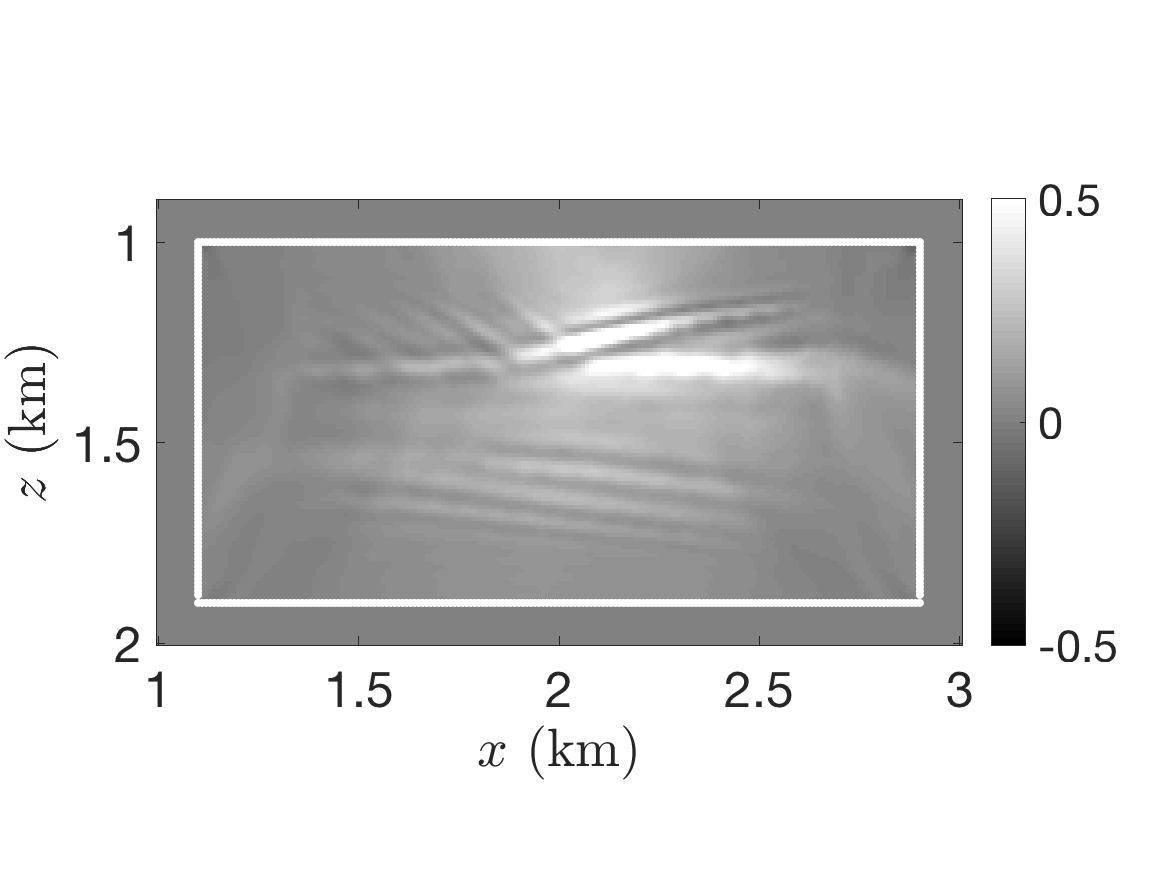}} 
 	\subfigure[]{\includegraphics[width = 0.4\textwidth,trim={0cm 1.5cm 0cm 3cm},clip]{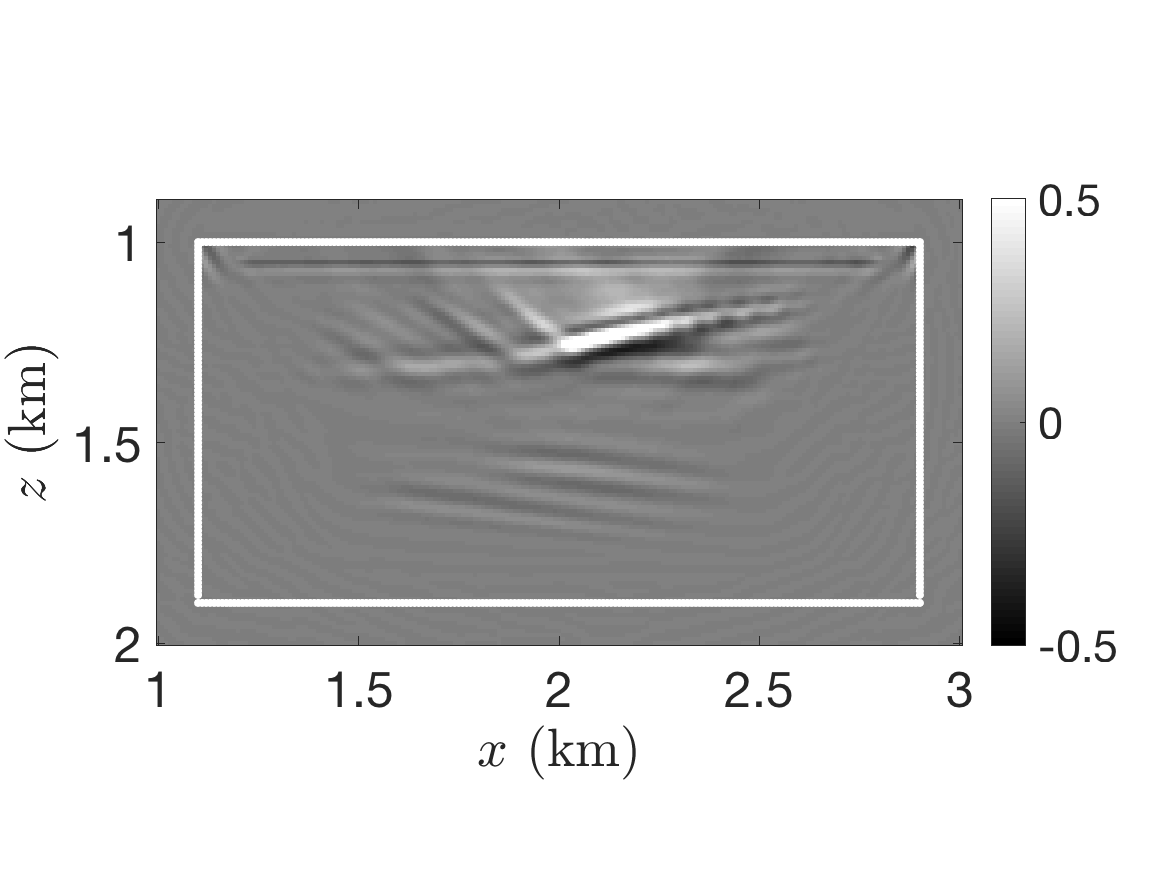}} 
  	\subfigure[]{\includegraphics[width = 0.4\textwidth,trim={0cm 1.5cm 0cm 3cm},clip]{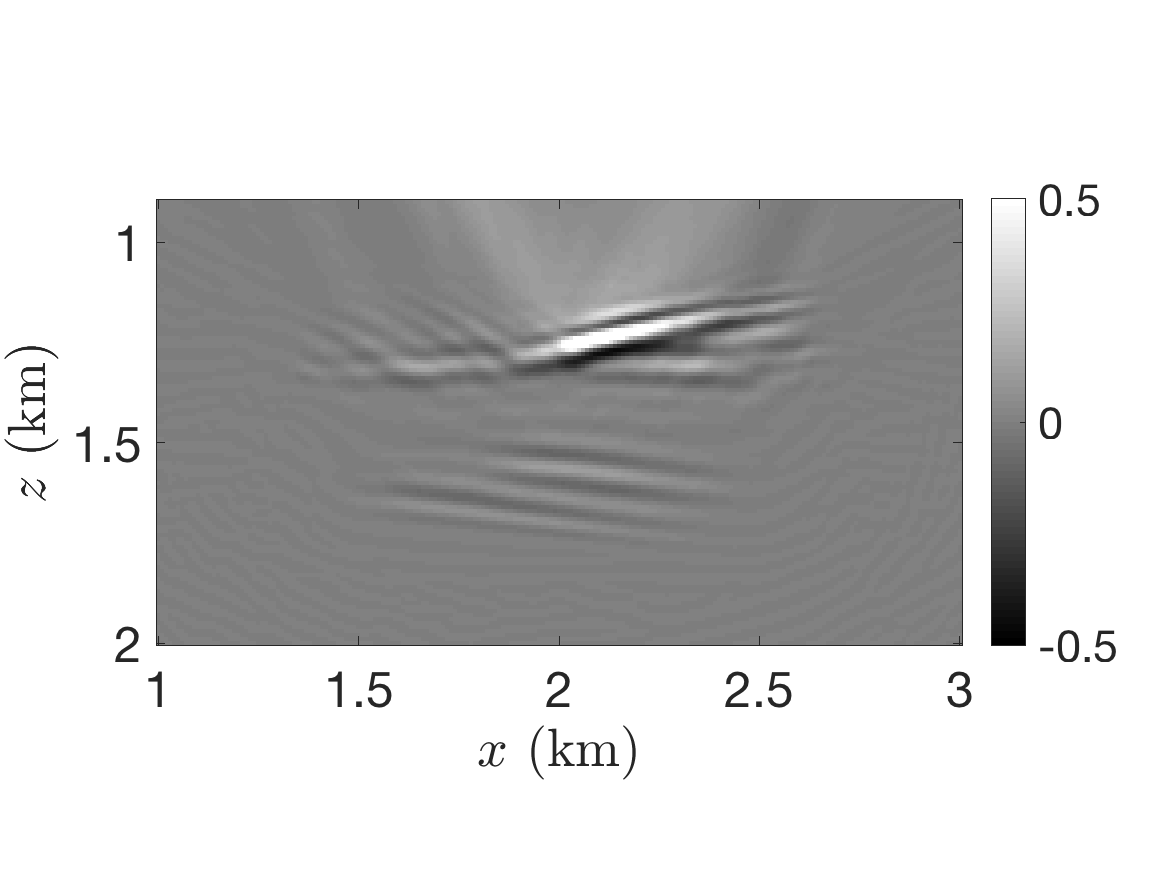}} 
 	\subfigure[]{\includegraphics[width = 0.4\textwidth,trim={0cm 1.5cm 0cm 3cm},clip]{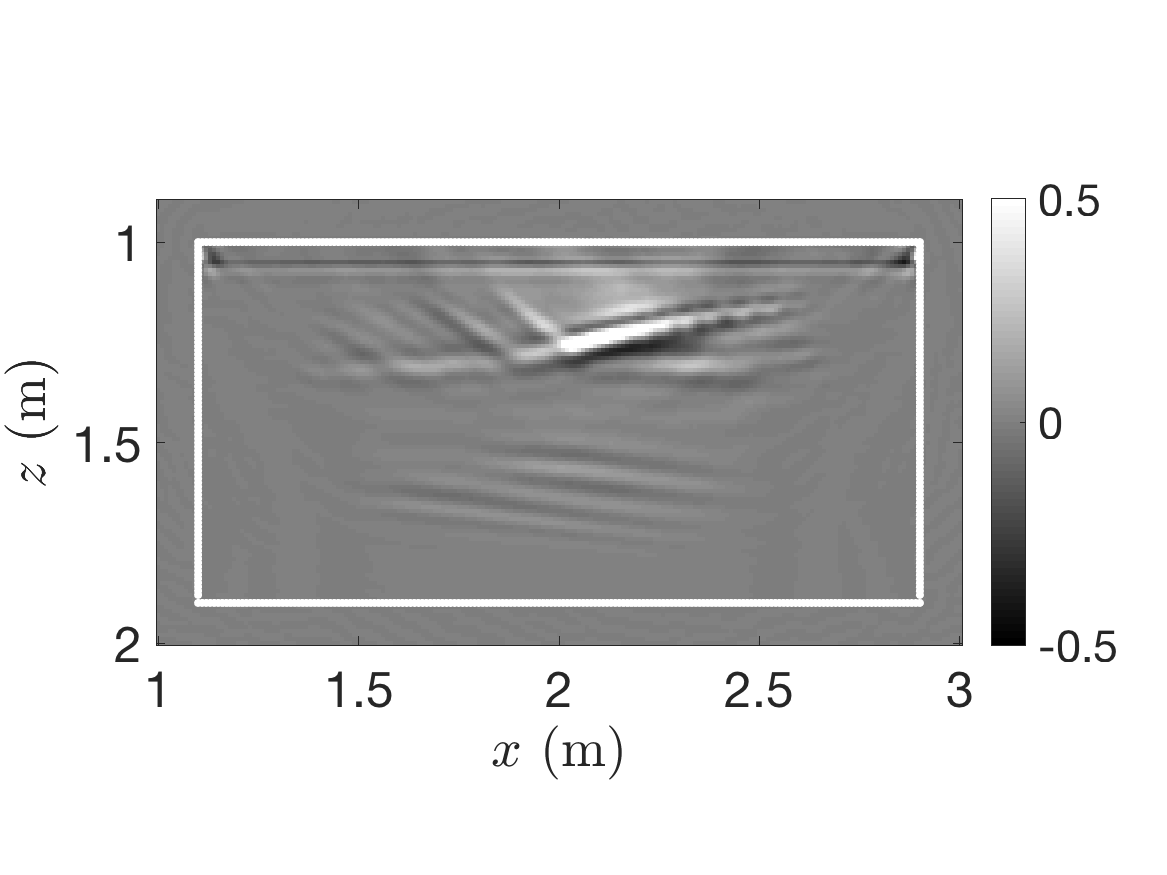}} 
	\caption{Gradient updates generated from five sources: (a) IFWI gradient update, (b) local FWI gradient update, (c) surface FWI gradient update, (d) local FWI gradient update from model with bottom reflector in Figure \ref{fig:true_model_refl}.}
 	\label{fig:gradients}
\end{figure}

\begin{figure} \centering
 	\subfigure[]{\includegraphics[width = 0.74\textwidth,trim={0cm 1cm 0cm 2.5cm},clip]{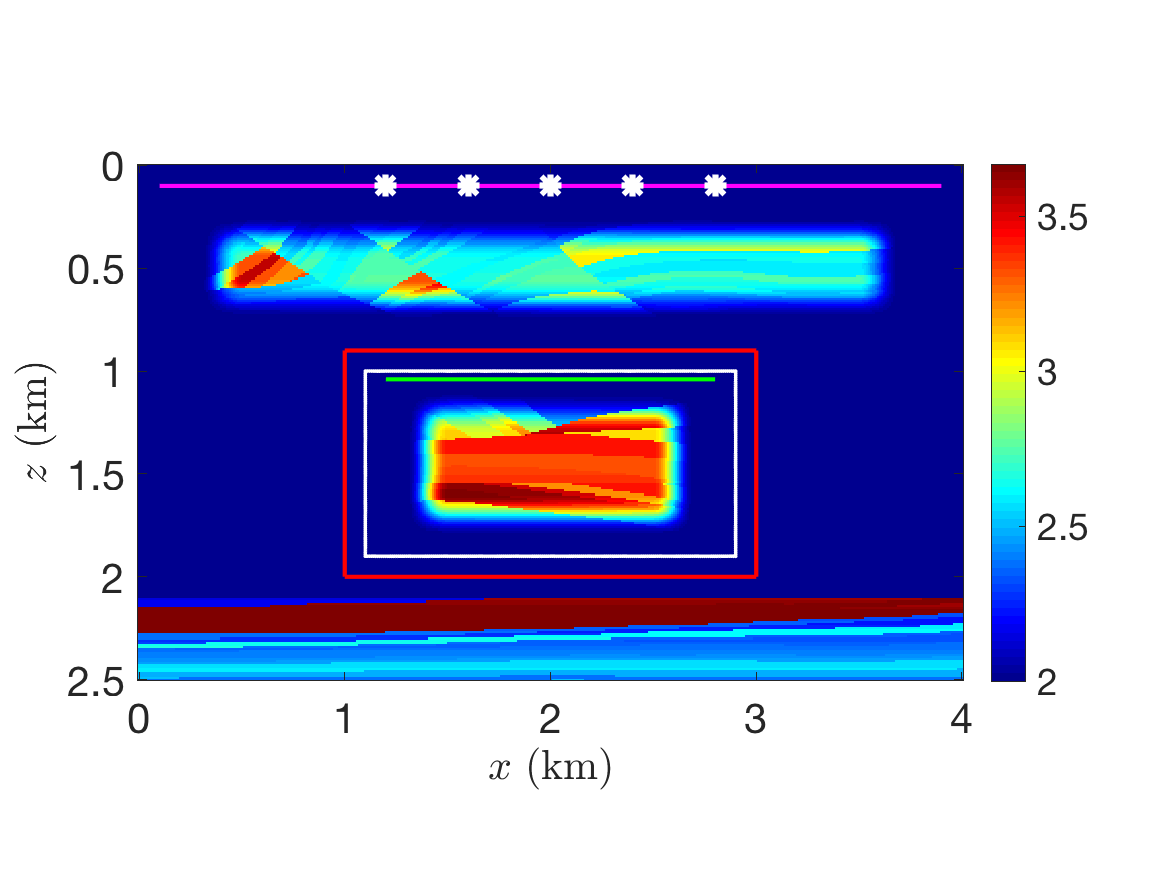}} 
	\caption{(a) True model with a reflector below the target. The white stars show physical surface sources, the magenta line represents physical surface receivers. The red line shows the local subdomain, the white line shows the injection boundary $\partial D$. Green line shows virtual receiver line for the convolution based local FWI method. }
 	\label{fig:true_model_refl}
\end{figure}

Figure \ref{fig:gradients} (a) shows the gradient update, computed in the local subdomain by IFWI using 5 equally spaced sources located at $z = 0.1$ km and $x = 1.2$ to 2.8 km in the initial model in Figure \ref{fig:true_model} (b). The update by the local FWI method of \cite{Cui:2020} is shown in Figure \ref{fig:gradients} (b) for comparison. It is interesting to note here that the local FWI method of \cite{Cui:2020} uses only the convolution field $p^{conv}$ and its adjoint computed from the misfit between the convolution field and redatumed data at the virtual receivers near the top boundary (green line in Figure \ref{fig:true_model}). It is impossible to generate low wavenumber updates with only a purely convolution field and its adjoint, if there are no up-going waves present in the forward convolution field. We observe that the update in Figure \ref{fig:true_model} (b) lacks the low wavenumbers. The conventional surface FWI update of the local domain is shown in Figure \ref{fig:gradients} (c) and has similar behaviour. In practice, there are reflecting horizons below the target that generate these up-going waves. For further insight into the presence of low-wavenumber gradient contributions, Figure \ref{fig:gradients} (d) shows the model update computed by the local FWI, where we added a reflector below the target in the true model, see Figure \ref{fig:true_model_refl}, and in this update the low wavenumbers are purposefully included by means of our model alteration. These contributions appear to improve stability and resolution of the convolution-based local FWI at least with exact data, as we show in the Examples section. Because of it intrinsic use of both reflected and transmitted fields irrespective of model parameters, the IFWI objective function does not require reflectors below the target to generate the low wavenumber components in the gradient, as both the up- and down-going waves are present in the correlation field. Nevertheless, presence of the up-going reflected waves in the convolution field seems to speed up the convergence of IFWI in the examples in the next section. 

We note also that even though the low wavenumber update appears possible with IFWI, a kinematically correct macro model is still important for redatuming, as an incorrect macro model of the target causes kinematic errors in the redatumed boundary data that propagate into the reconstruction. We investigate this issue in the Examples section with a proxy example. 
 
\subsection{Computational cost}

In this section we express the computational cost of the proposed method and compare it with that of the other FWI methods used in this paper as comparison. First, we consider the cost of IFWI. We use capital letters for full domain and surface quantities, and small letters for local subdomain quantities.

In the implementation of IFWI, the computational cost is dominated by the forward and adjoint modelling. The cost of one constant-density vector-acoustic PDE solve in flops is proportional to $n_x n_z n_t$ in 2D and to $n_x n_y n_z n_t$ in 3D, where $n_x$, $n_y$ and $n_z$ are the dimensions of the local subdomain in grid-points and $n_t$ is the number of time steps. A total of four PDE solves on the local subdomain are required per source for one IFWI gradient evaluation: two forward solves, equations (\ref{conv_constr}) and (\ref{corr_constr}), and two adjoint solves, equations (\ref{adj_conv}) and (\ref{adj_corr}). Thus, the overall cost of IFWI per gradient calculation is on the order of $\sim 4 C n_x n_z n_t N_s$ in 2D and $\sim 4 C n_x n_y n_z n_t N_s$ in 3D, where $N_s$ is the number of surface sources and the proportionality constant $C$ depends on the finite difference scheme and the size of the PML layers. Typically, the cost for the PML calculation is per one grid point is higher, as the FD equations include extra terms involving the damping functions. Based on the size of the local domain that we use in the examples, the 4th order accurate discretization in space and 2nd order accurate discretization in time, and PML width of 20 points, $C\approx 100$ for our current 2D implementation. This constant can be reduced at the expense of the code clarity and a small increase in memory use. Assuming that $n_x, n_y, n_z \sim n$, the cost of one iteration of IFWI is $\sim O(n^2 n_t N_s)$ in 2D and $\sim O(n^3 n_t N_s)$ in 3D.

The cost of redatuming, which only needs to be performed once, must be added to the cost of waveform inversion. Limiting ourselves here to data-driven redatuming, and more specifically Marchenko redatuming by iterative substitution \citep{Neut:2015, Ravasi:2017}, the overall cost is dominated by evaluation of multi-dimensional convolutional operators. More specifically, a multi-dimensional convolutional operator involves a fast Fourier transform (FFT), a batched matrix-vector multiplication and an inverse FFT. Two such multi-dimensional convolutions are required per iteration of redatuming. Moreover, two applications of a muting operator are also necessary at each iteration, see equations 29 and 30 of \cite{Neut:2015}. The number of iterations required for the Marchenko equations to converge is typically small, $O(10)$. Based on the cost of each operation, we estimate the overall cost of redatuming to be on the order of:
\begin{eqnarray}
	\mbox{Cost (Red)} &=& C_{red} \; n_{red} \; n_{iter} 2 [(2 \; n_t\; log(n_t)\;  N_r 
	+ n_t \; (4 \;N_s\;(2\;N_r-1))) +  n_t \;N_r] \label{redatum_cost}
\end{eqnarray}
where
\begin{description}
 	\item $n_{red}$ is the number of points in the subsurface where we need to redatum the fields
 	\item $n_{iter}$ is the number of iterations of the Marchenko equations per one point
	\item $N_r$ is the number of surface receivers
	\item $C_{red}$ is an unknown proportionality constant.
\end{description} 
For the terms in the square brackets, the first corresponds to forward and inverse FFTs, the second term is the discretized integral over the acquisition surface in complex-number arithmetic, and the third term is the application of muting. The factor of 2 in front of the square brackets arises from the fact that two multi-dimensional convolutions are required per iteration of redatuming. Finally, $n_{red} \sim 4 (n_x + n_z)$ in 2D and $n_{red} \sim 4 (n_x n_z + n_x   n_y + n_y  n_z)$ in 3D. This accounts for two horizontal and two vertical subdomain boundaries (four in 3D), to each of which pressure measurements need to be redatumed at two layers of points in order to obtain displacement components/pressure derivatives \citep{Cui:2020}.

In order to compare the computational cost of local FWI and redatuming, we observe that the cost of redatuming is dominated by $n_t$ and $N_r$. We assume $N_r$ to be in the order of $O(N_x)$ in 2D and $O(N_x N_y)$ in 3D, where $N_x$, $N_y$ represent the dimensions of the full domain in grid-points. Moreover, if $n_x, n_y, n_z \sim n$,  and $N_x, N_y \sim N$ then redatuming the surface data to the whole local domain boundary has complexity $\sim O(n N n_t N_s)$ in 2D, which is comparable to the order of complexity of one iteration of IFWI, and in practice is likely to have the cost of a few iterations of IFWI due to other multiplicative factors in equation \ref{redatum_cost} and the fact that $N>n$. In 3D, the cost of redatuming is $O(n^2 N^2 n_t n_s)$, which is larger by a factor of $N$ than one iteration of FWI.

Conventional surface FWI requires two PDE solves on the full domain per source: one forward and one adjoint solve, and has the cost $\sim O(N^2 n_t N_s)$ in 2D and $\sim O(N^3 n_t N_s)$ in 3D per iteration. If $V_{full} = N_x (N_y) N_z$ and $V_{local} = n_x (n_y) n_z$ denote the volume of the full model and the volume of the local subdomain in grid-nodes, and they are discretized using the same temporal grid, then the cost of IFWI roughly becomes:
\[
	\mbox{Cost}(\mbox{IFWI}) = 2 \frac{V_{local}}{V_{full}} \mbox{Cost}(\mbox{surface FWI}) + \mbox{Cost(Redatuming)}.
\]

Therefore, the IFWI method is practical in 2D compared to the conventional surface FWI when the local domain occupies less than roughly 1/2 of the full model. Due to the cost of redatuming in 3D, 1/2 reduction in the domain size might not be sufficient to offset the redatuming cost. In local domain applications the goal is to make $\frac{V_{local}}{V_{full}}$ as small as possible. In the proposed implementation, since the side injection boundaries are included in the forward modelling process, the local subdomain size can be reduced both in width and height.

When compared to the local convolution-based FWI of \cite{Cui:2020}, IFWI is admittedly twice as expensive per gradient evaluation for the same local domain size, number of sources and discretization, since local FWI requires only two PDE solves per source per gradient evaluation. Local FWI requires additional $O(n_x)$ redatuming steps in 2D and $O(n_x n_y)$ redatuming steps in 3D to redatum pressure to the virtual receivers. In the examples section, we revisit the cost comparison of IFWI and local convolution-based FWI methods, taking into account additional consideration arising from reconstruction quality.

The cost estimates for the three inversion methods are summarized in Table \ref{cost}.

\begin{table*}
 \caption{Computational cost of IFWI compared to surface FWI and local convolution-based FWI.}
 \label{cost}
 \begin{tabular}{@{}lcccc}
 	\hline
  	Method 		& Inversion cost 	& Inversion cost  	& Redatuming cost 	& Redatuming cost  \\
				& per iteration, 2D	& per iteration, 3D	& 2D				& 3D \\
	\hline
   	IFWI 		& $O(n^2 n_t N_s)$ 	& $O(n^3 n_t N_s)$ 	& $O(n N n_t N_s)$ 	& $O(n^2 N^2 n_t n_s)$\\
	Surface FWI 	& $O(N^2 n_t N_s)$ 	& $O(N^3 n_t N_s)$ 	& $-$ 			& $-$ \\ 
	Local FWI 	& $O(n^2 n_t N_s)$ & $O(n^3 n_t N_s)$ &  $O(n N n_t N_s)$ & $O(n^2 N^2 n_t n_s)$\\
 	\hline
	\end{tabular}
\medskip \\
\end{table*}

\section{Examples}

\subsection{Exact redatuming}

In this section, we demonstrate the performance of the interferometric FWI on stylized examples with exact redatuming. We compare performance of our method to the performance of surface FWI and the local convolution-based FWI of \citep{Cui:2020} (local FWI). For the first example, the true velocity model is the middle part of the velocity model shown in Figure \ref{fig:true_model} (a) between $x=1$ and $x=3$ km to reduce the cost of full domain modelling and inversion.

For the second example, we add a reflector below the target. The true model with reflector is shown in Figure \ref{fig:true_model_refl}, we also use the middle part of it between $x=1$ and $x=3$ km. 

We place five evenly spaced sources at depth $z = 0.1$ km from $x = 0.2$ to $x = 1.8$ km. For the conventional FWI, we use the receivers at a depth of $z = 0.1$ km from $x = 0.1$ to $x = 1.9$ km. As before, the local domain is marked by the red line, the white line denotes the injection boundary $\partial D$, and the green line denotes the receivers used in the local FWI method.

For the proposed IFWI method and the local FWI, we start with the initial model shown in Figure \ref{fig:true_model} (c). For surface FWI, we additionally apply smoothing by Gaussian filter with STD of 20 m and 80 m to the overburden and the bottom reflector.

For the local FWI we do not apply any preconditioning, while for FWI we suppress the receiver footprint during the inversion, since in practice, receivers are likely to be in the water. We note that suppressing the receiver footprint  for local FWI does not change the result significantly. 
We stop all inversions after 200 iterations, regardless of whether any in-built stopping criterion of the L-BFGS solver was reached.

\subsubsection{Example 1: exact redatuming, no bottom reflector}

\begin{figure}\centering
 	\subfigure[]{\includegraphics[width = 0.45\textwidth,trim={0cm 1.5cm 0cm 3cm},clip]{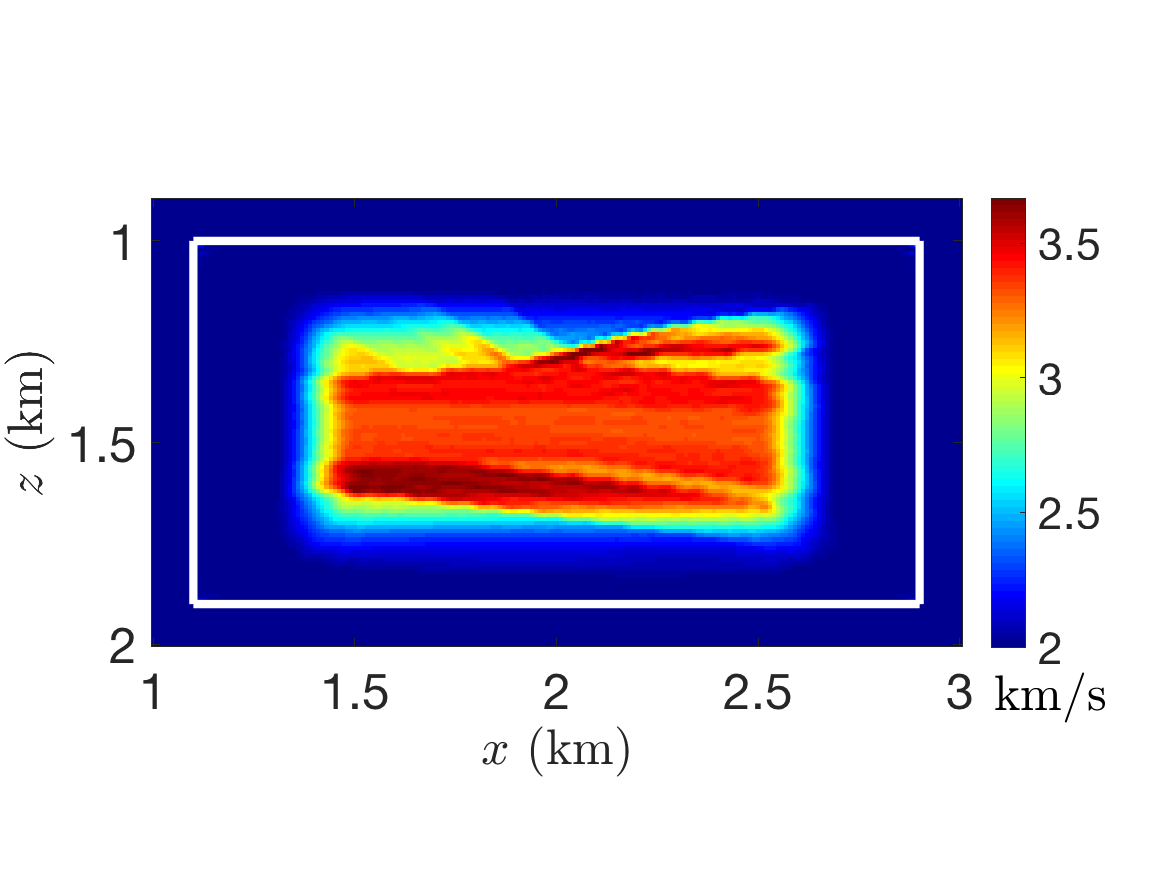}} 
 	\subfigure[]{\includegraphics[width = 0.45\textwidth,trim={0cm 1.5cm 0cm 3cm},clip]{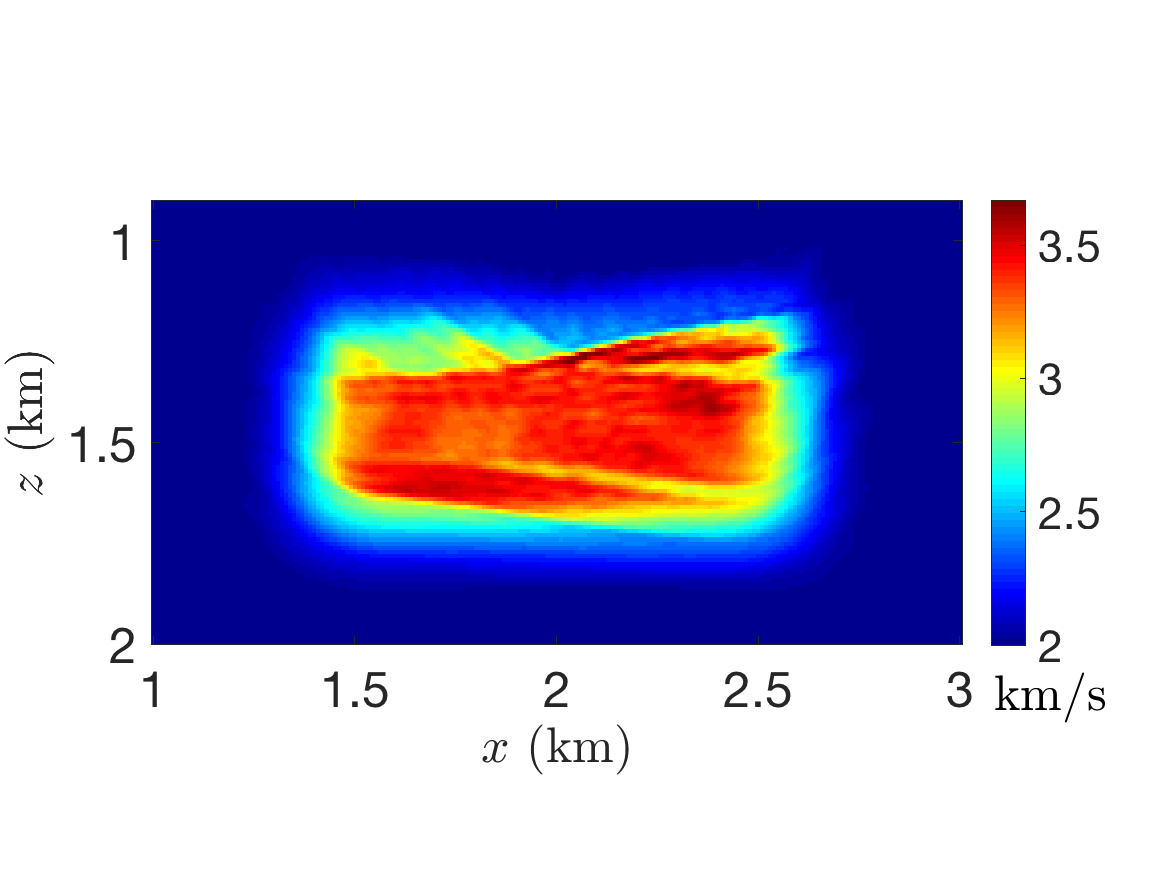}} 
 	\subfigure[]{\includegraphics[width = 0.45\textwidth,trim={0cm 1.5cm 0cm 3cm},clip]{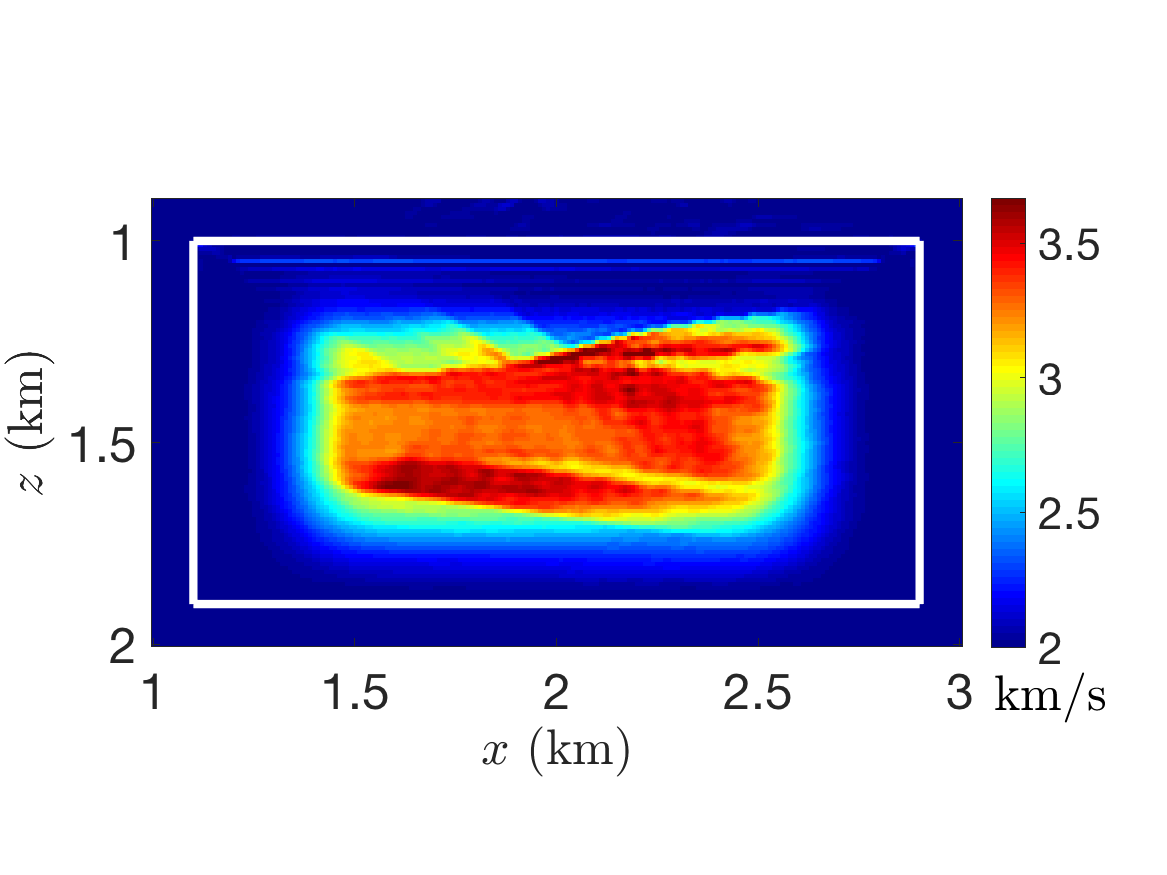}} 
	 \subfigure[]{\includegraphics[width = 0.45\textwidth,trim={0cm 1.5cm 0cm 3cm},clip]{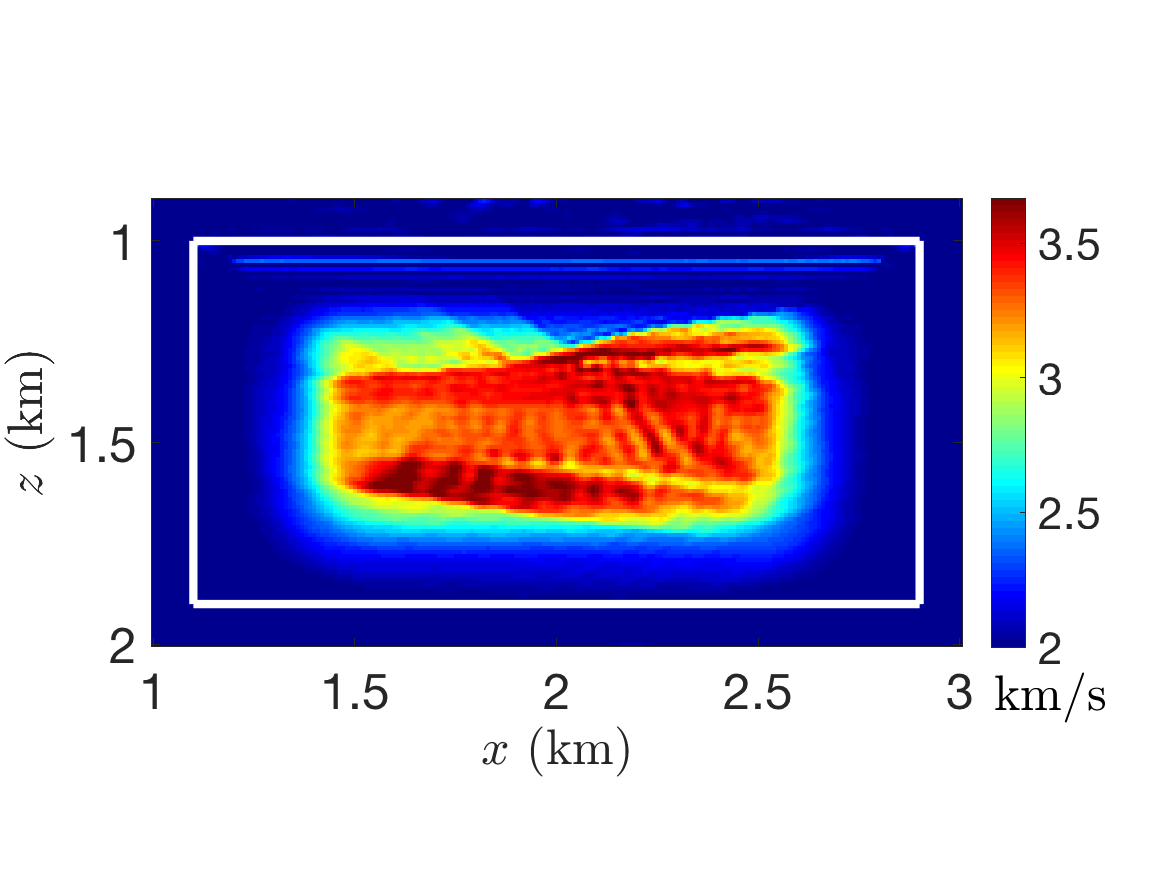}} 
	\caption{FWI reconstructions of the target area: (a) IFWI atfer 200 iterations, (b) conventional surface data FWI after 200 iterations, (c), (d) local convolution-based FWI after 100 and 200 iterations. Only the local part is shown for the conventional FWI in image (b).}
 	\label{fig:Ex1_recoveries}
\end{figure}

Figure \ref{fig:Ex1_recoveries} (a) and (b) show the reconstruction of the target obtained by interferometric FWI and conventional FWI, respectively. Similarly, Figure \ref{fig:Ex1_recoveries} (c) and (d) show the reconstruction of the target obtained by the local FWI after 100 and 200 iterations, respectively. For the conventional FWI we only show the target area in Figure \ref {fig:Ex1_recoveries} (b). Overall, the interferometric FWI accurately recovers all features of the target area, including the velocity and shape of the lowest part of the target. Both the conventional and local FWI accurately image the top part of the target, however the velocity of the bottom high velocity feature layer is underestimated, and both inversions produce rounded shape at the bottom corners of the target. There are also variations in the velocity in the middle layer of the target that are not present in the true model.  We also note that the local FWI has developed artifacts in the form of high frequency noise at iteration 200, while the reconstruction is more accurate at iteration 100. These artifacts may be indicative of data overfitting.

\begin{figure}\centering
 	\subfigure[]{\includegraphics[width = 0.45\textwidth,trim={0cm 0cm 0cm 0cm},clip]{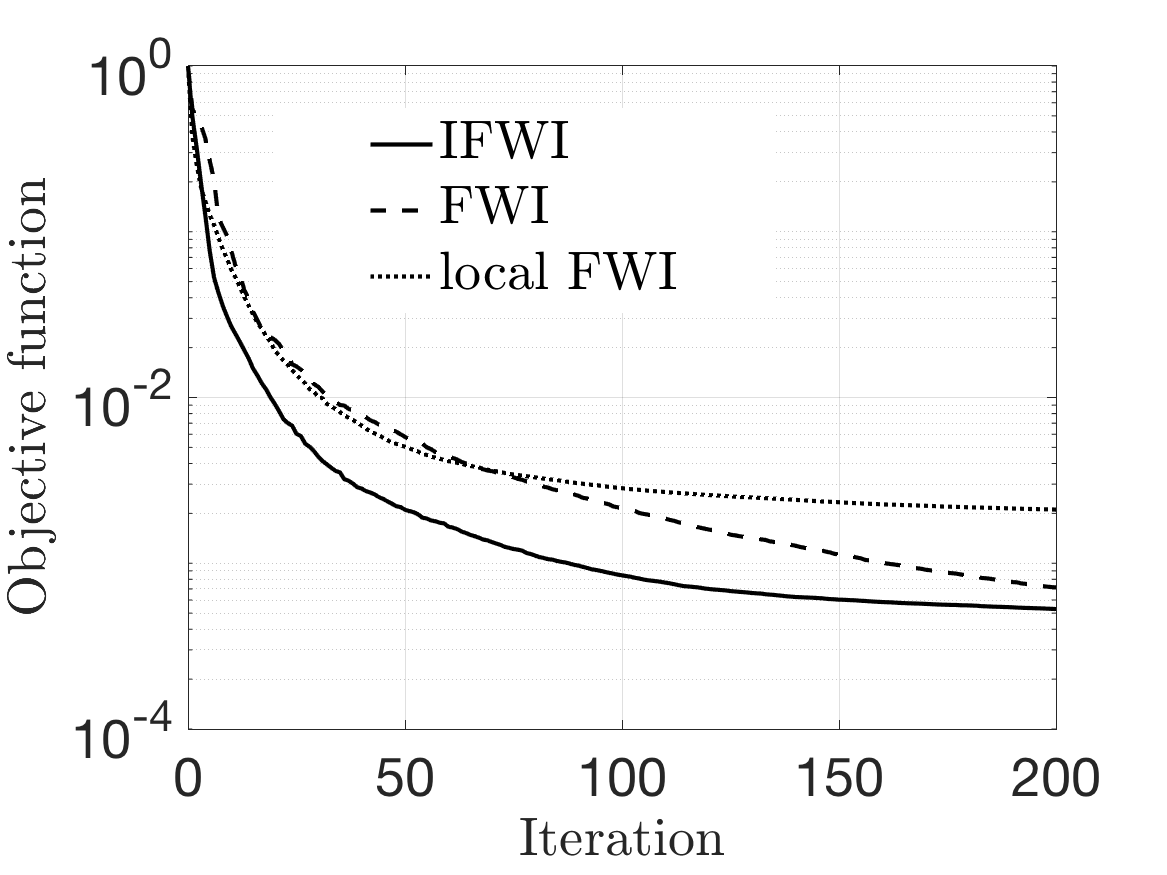}} 
 	\subfigure[]{\includegraphics[width = 0.45\textwidth,trim={0cm 0cm 0cm 0cm},clip]{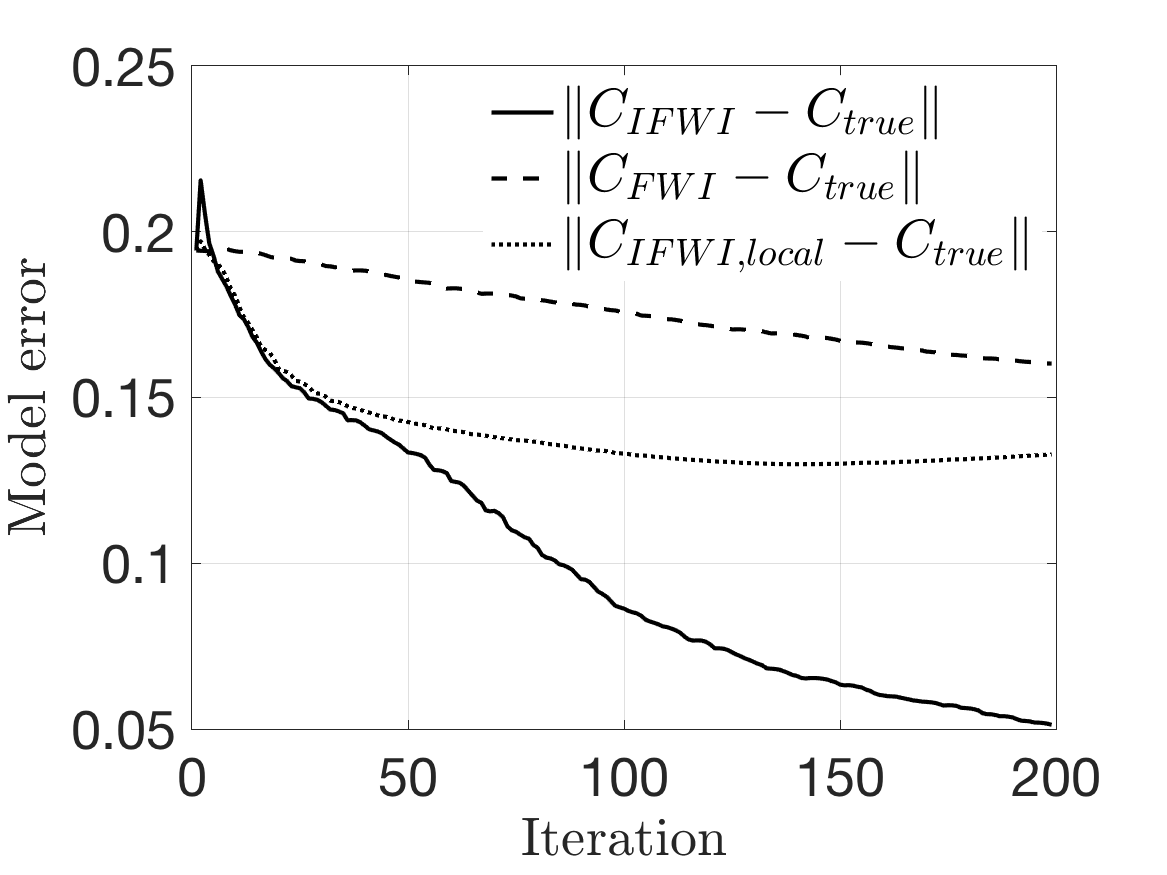}} 
	\caption{(a) Objective function with iteration for the three inversions, (b) $L_2$ norm of the model error in the local domain.}
 	\label{fig:Ex1_errors}
\end{figure}

Figure \ref{fig:Ex1_errors} (a) and (b) shows the objective function and the $L_2$ norm of the model error in the local domain, where the model error norm is related to the inner product (equation \ref{model_ip}). We observe faster convergence for the interferometric inversion than for both other methods. For the local FWI the convergence slows down after about 100 iterations, which is indicative that the method has converged. After that the local FWI objective function is minimally reduced, however, the root-mean-square model error begins to grow after $\sim$140 iterations. For the other two methods, the objective function and the model RMS error both decrease until the end of the inversion run. It is notable that for after about 130 iterations, the IFWI model residual decreases faster than its objective function. Running FWI and IFWI to 200 iterations leads to visible model improvement. The final model RMS error is significantly smaller for IFWI than for both other methods.

\begin{figure}\centering
 	\subfigure[]{\includegraphics[width = 0.45\textwidth,trim={0cm 1.5cm 0cm 3cm},clip]{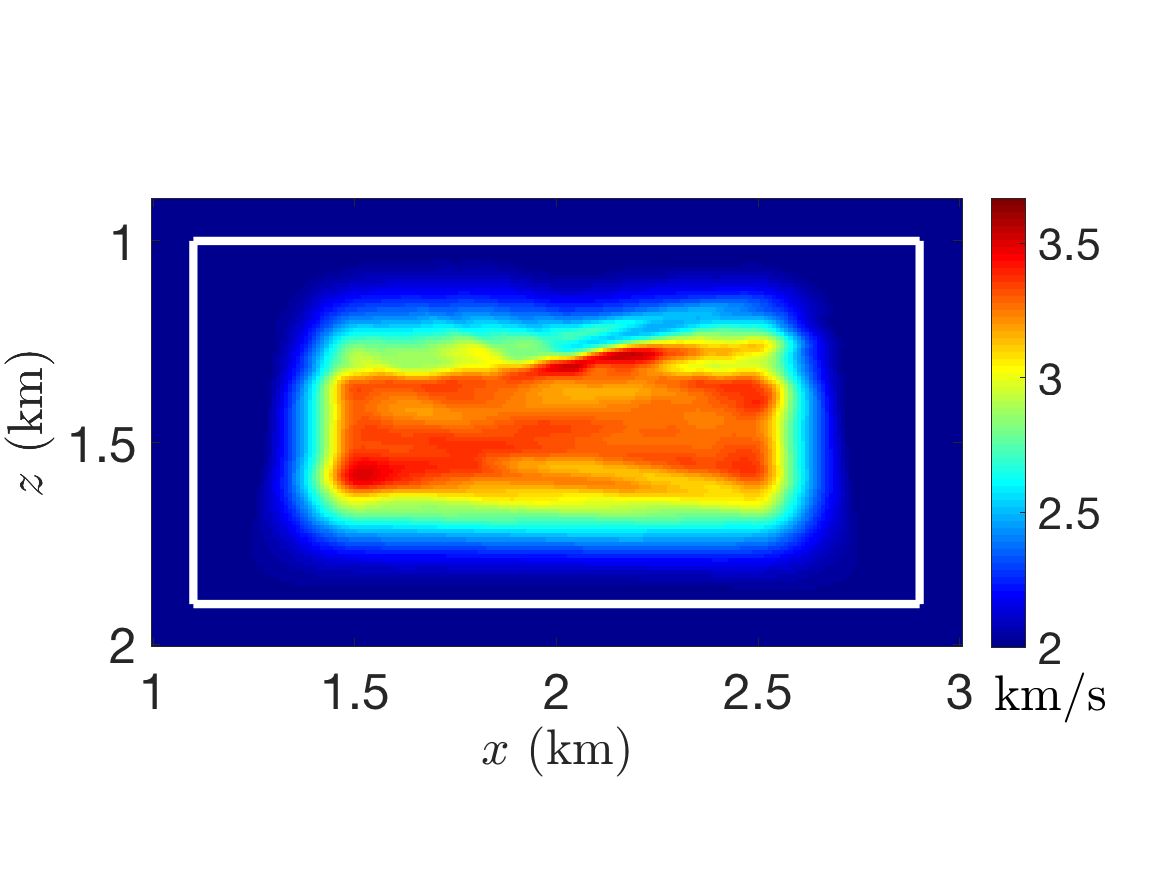}} 
 	\subfigure[]{\includegraphics[width = 0.45\textwidth,trim={0cm 1.5cm 0cm 3cm},clip]{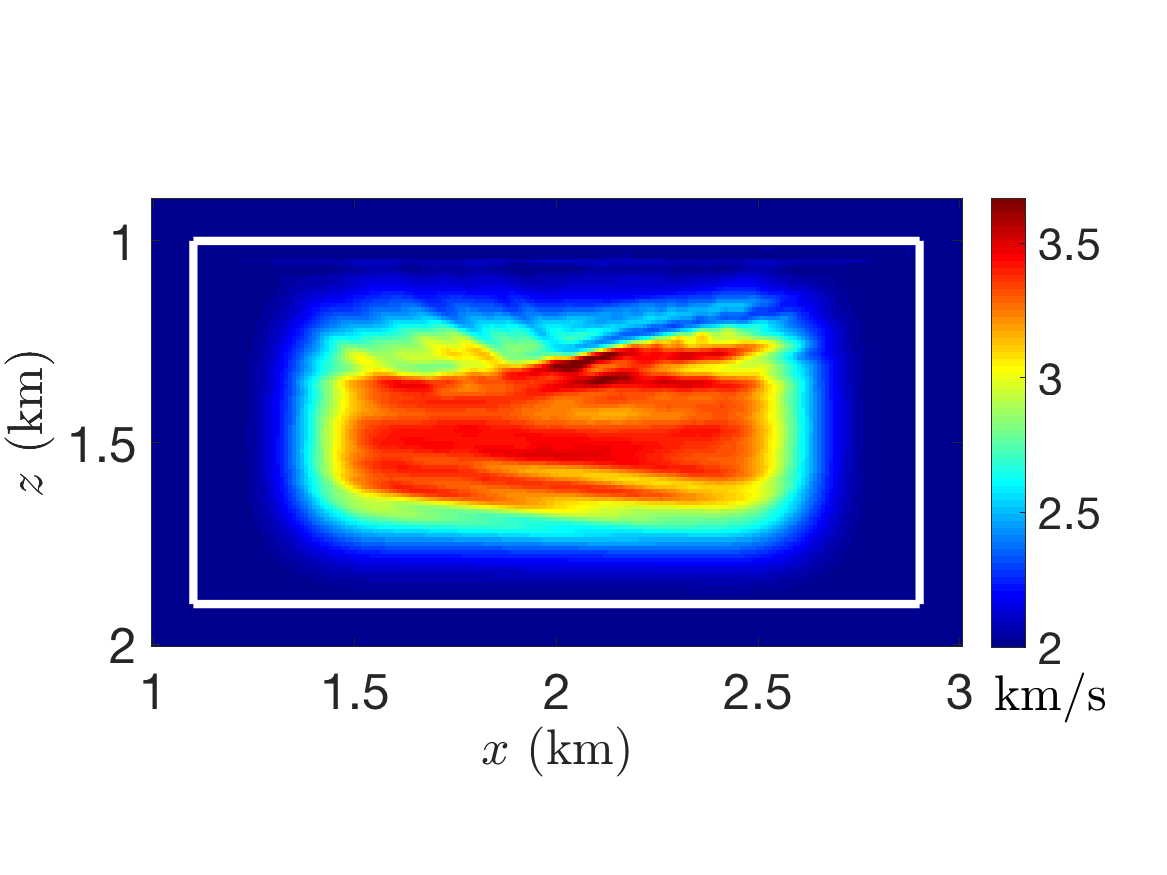}} 
	\caption{Example 1, reconstructions of the target area: (a) IFWI, at iteration 5, (b) local FWI, at iteration 5. }
 	\label{fig:Ex1_iterations5_20}
\end{figure}

Figure \ref{fig:Ex1_iterations5_20} shows the inversion progress of IFWI and the local FWI at iteration 5. It is apparent from this Figure that the interferometric and local inversions proceed along different update paths. At the initial iterations, local FWI updates significantly more high wavenumbers in the model, whereas the interferometric FWI updates mostly the macro velocity model, particularly, the corners of the target. 

\subsubsection{Example 2: exact redatuming, reflector below the target}

\begin{figure} \centering
 	\subfigure[]{\includegraphics[width = 0.45\textwidth,trim={0cm 1.5cm 0cm 3cm},clip]{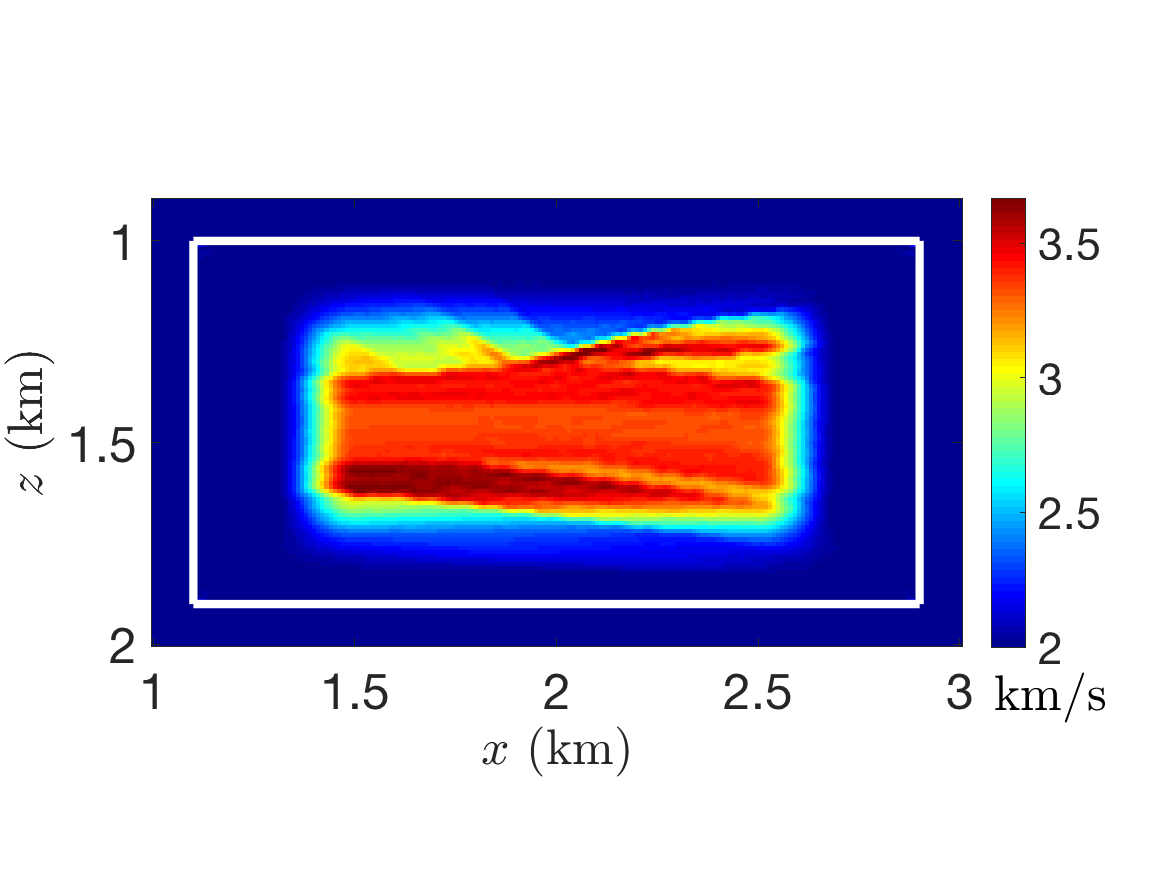}} 
 	\subfigure[]{\includegraphics[width = 0.45\textwidth,trim={0cm 1.5cm 0cm 3cm},clip]{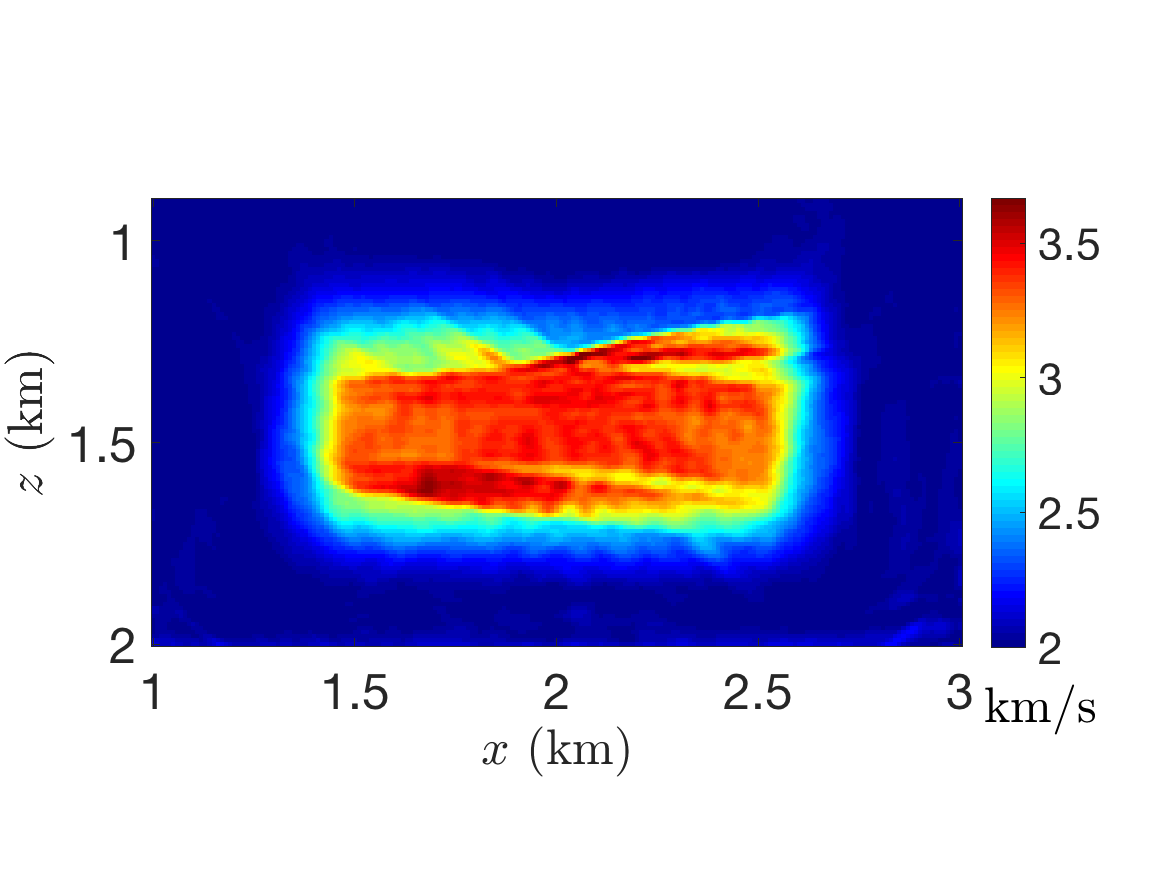}} 
 	\subfigure[]{\includegraphics[width = 0.45\textwidth,trim={0cm 1.5cm 0cm 3cm},clip]{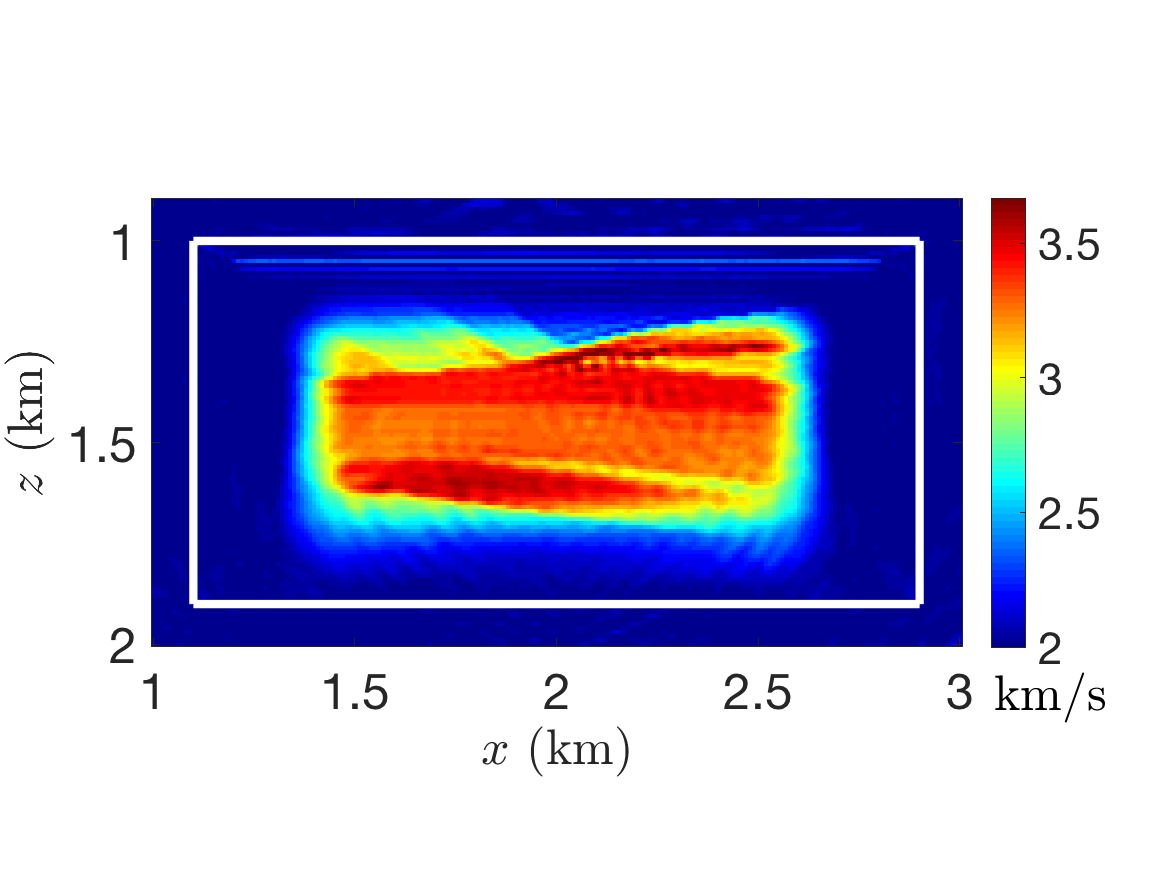}} 
	\subfigure[]{\includegraphics[width = 0.45\textwidth,trim={0cm 1.5cm 0cm 3cm},clip]{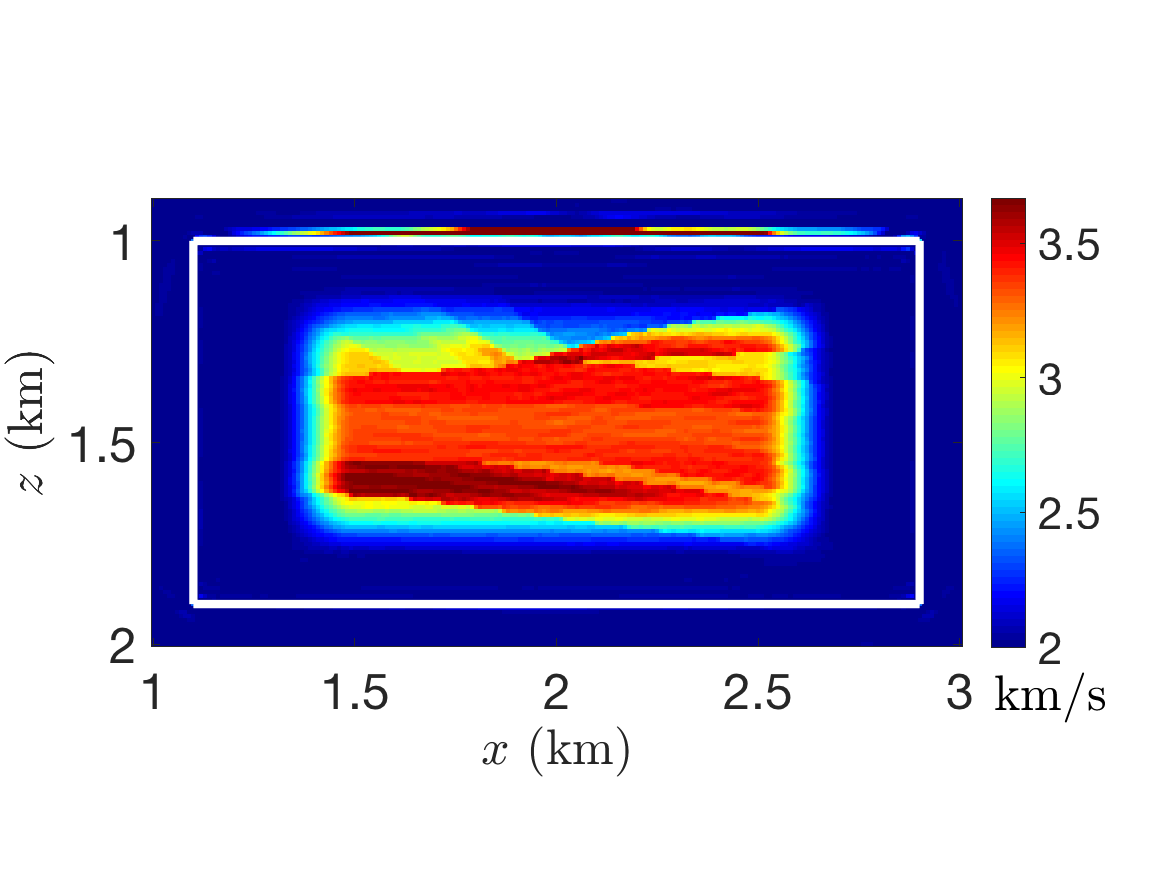}} 
	\caption{Example 2, true model with the bottom reflector, reconstructions. (a) Interferometric FWI after 100 iterations; (b) surface data FWI after 200 iterations; (c) local convolution-based FWI after 200 iterations; (d) local convolution-based FWI with the misfit calculated at every point in the local subdomain after 200 iterations of LBFGS.}
 	\label{fig:Ex2_init_recoveries}
\end{figure}

\begin{figure}\centering
 	\subfigure[]{\includegraphics[width = 0.45\textwidth,trim={0cm 0cm 0cm 0cm},clip]{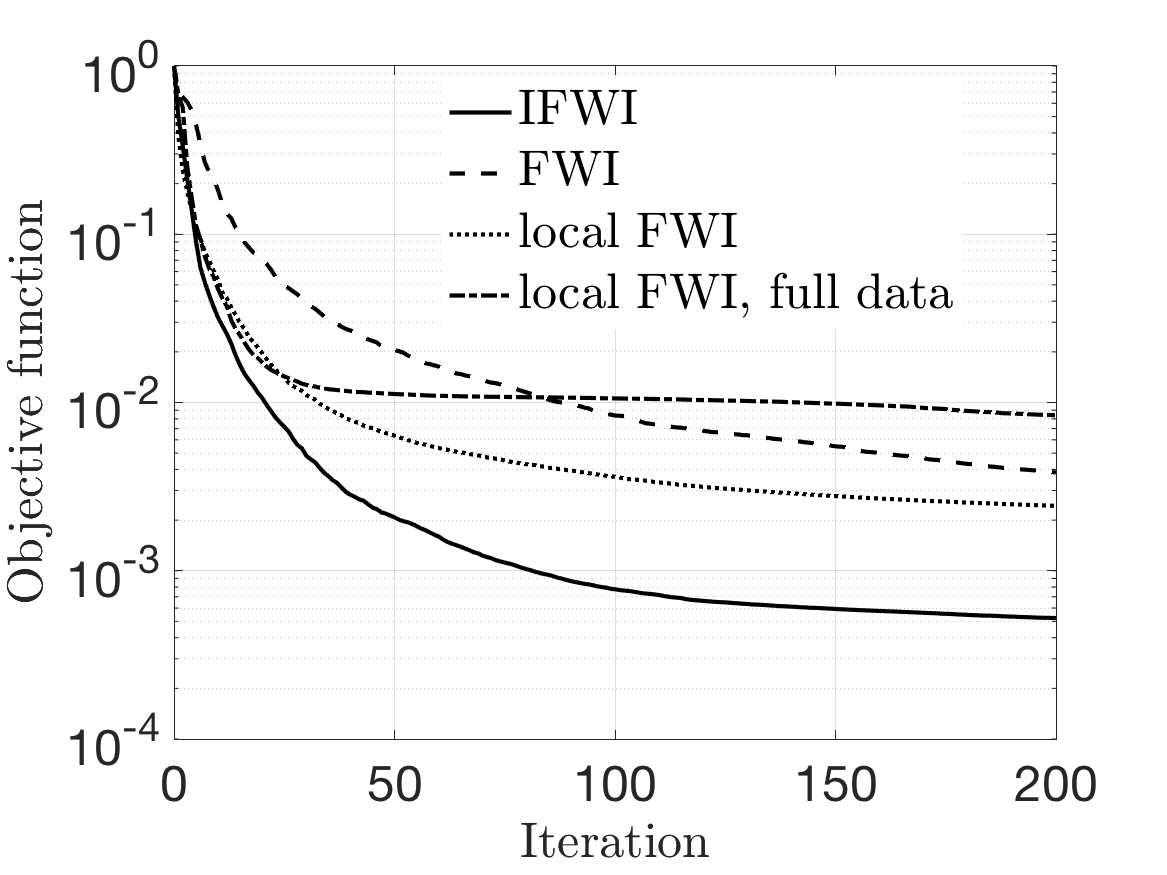}} 
 	\subfigure[]{\includegraphics[width = 0.45\textwidth,trim={0cm 0cm 0cm 0cm},clip]{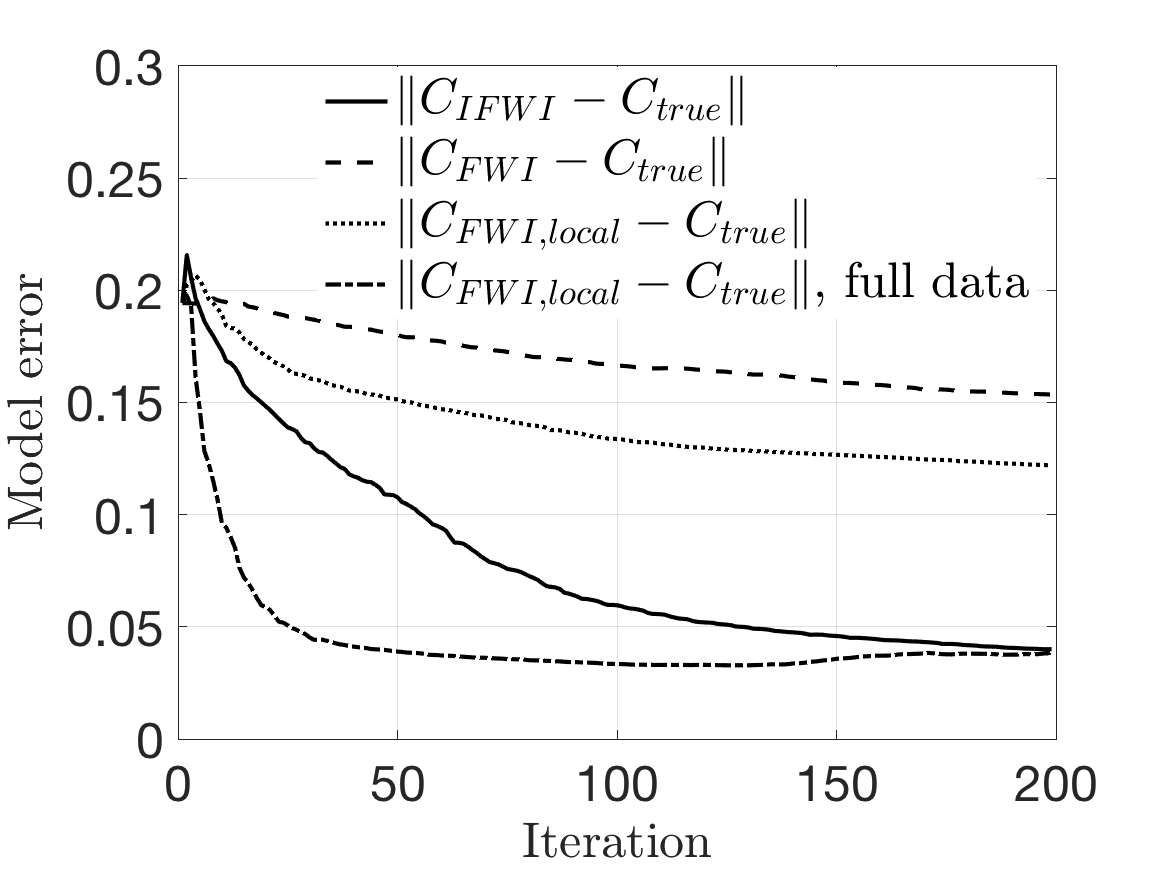}} 
	\caption{(a) Objective function with iteration for the three inversions from the true model with the bottom reflector, (b) $L_2$ norm of the model error in the local domain.}
 	\label{fig:Ex2_errors}
\end{figure}

Figure \ref{fig:Ex2_init_recoveries} shows the reconstruction of the target obtained by IFWI (a), conventional FWI (b) and local FWI (c) after 100, 200 and 200 iterations respectively. Figure \ref{fig:Ex2_errors} shows the objective function (a) and the RMS model error (b) for this example. The reconstruction for the local FWI is much more stable than in the previous example, the model RMS error decreases and the reconstruction visually improves until the end of the inversion run. The IFWI converges faster and reaches the same RMS error as in the previous example after $\sim 130$ iterations and produces a higher resolution image in both examples than the two other methods.

\cite{Cui:2020} point out that their method in not constrained by practical considerations to have redatumed data at virtual receivers only at the top of the local subdomain, and other locations can also be used. In the proposed IFWI method, we measure the misfit in the objective function everywhere in the local subdomain. The same can be done with the method of \cite{Cui:2020}. As we show below, such misfit measure significantly improves convergence and resolution of that method, but also dramatically raises the redatuming cost compared to the proposed IFWI method. 

In Figure \ref{fig:Ex2_init_recoveries} (d), we show the  inversion result obtained by the method of \cite{Cui:2020} where the misfit is also computed everywhere inside the local subdomain just as for our proposed IFWI method. 
Figure \ref{fig:Ex2_errors} shows the objective function and the RMS model error plot for this inversion as dash-dot lines. With the data measurements everywhere, the local FWI converges in about 50 iterations to a model comparable in resolution to the IFWI recovery achieved after about 150 iterations. This is not surprising, since with full data available everywhere in the local subdomain, the convolution-based inverse problem becomes the problem of solving an overdetermined linear system
\[
	(\Lb \wb) m = \s
\] 
where $\Lb \wb$ and $\s$ are known everywhere. So it is a linear inverse problem, while IFWI remains a non-linear problem. Stated differently, the local FWI compares the modelled convolution field to the "known true" field everywhere, while IFWI compares two modelled fields to each other without access to the "true field". 

Clearly, the ability to evaluate the objective function at every point in the local subdomain has a substantial impact on image resolution. However, the need to have pressure measurements at every point inside the local subdomain for the convolution-based local FWI of \cite{Cui:2020} raises the cost of redatuming for that method from $O(n N n_t N_s)$ to $O(n^2 N n_t n_s)$ in 2D and from $O(n^2 N^2 n_t n_s)$ to $O(n^3 N^2 n_t n_s)$ in 3D. While doing this is still possible, this extra redatuming cost is likely to offset in 2D and exceed in 3D the extra cost per iteration and slower convergence of the proposed IFWI method. On the other hand, in the IFWI method, the correlation forward and adjoint fields are obtained at the cost of two additional local domain modelings per iteration. Thus, the same result is obtained by IFWI at a similar or lower cost.

\subsection{Inexact and missing data}

\subsubsection{Example 3: inexact redatuming} \label{incorrect data}

\begin{figure}\centering
	\subfigure[]{\includegraphics[width = 0.45\textwidth,trim={0cm 1.5cm 0cm 3cm},clip]{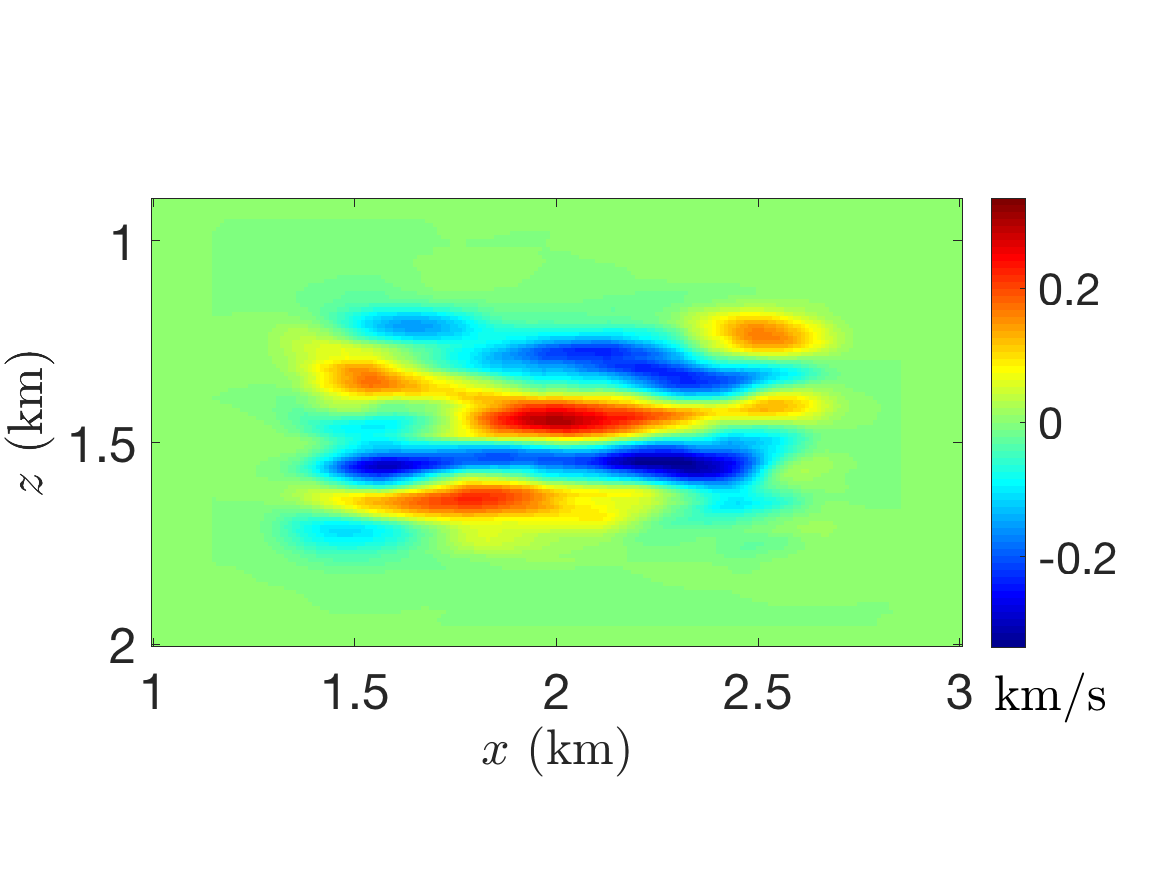}} 
	\subfigure[]{\includegraphics[width = 0.45\textwidth,trim={0cm 1.5cm 0cm 3cm},clip]{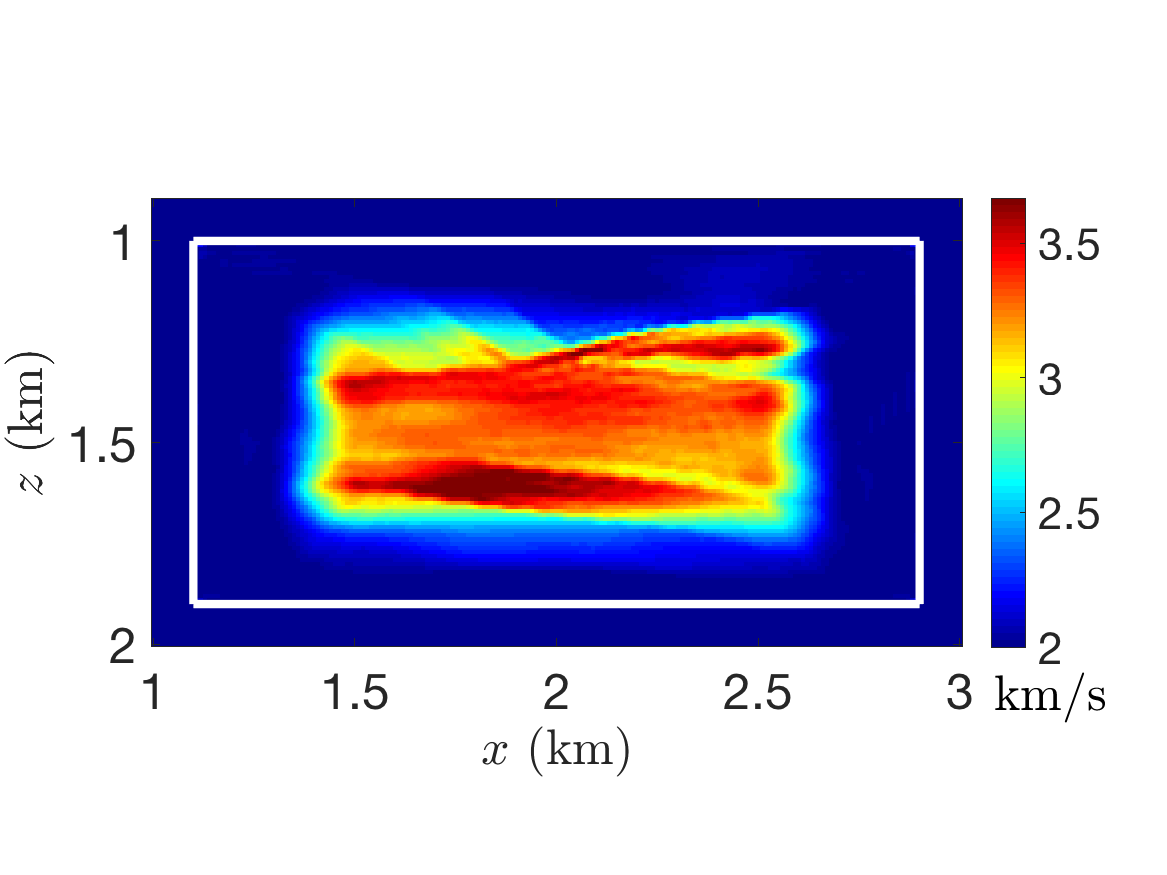}} 
	\caption{(a) Random bias added to the initial model, and the true model in which the side and bottom boundary data is generated; (b) recovery by IFWI of biased data from the biased initial model. }
 	\label{fig:Ex2_rand_bias_init_recoveries}
\end{figure}

As mentioned above, the kinematic errors in the macro velocity model can lead to incorrect arrival times of the events in the redatumed fields. The local inversion can not fix the kinematic errors in the overburden. However, due to the low wavenumber components in the IFWI gradient, it is still possible to correct kinematic errors in the target area, if the redatumed data are themselves correct. At the same time, the incorrect kinematics inside the local domain also affect redatuming, introducing inaccuracies in the redatumed data. To get a feeling for the sensitivity of the proposed method to data generated in a kinematically incorrect macro velocity model and the ability of the method to cure incorrect kinematics  of the target area during the inversion, we make a biased model where we introduce smooth random velocity variations in the target area of both the true and initial velocity models. The velocity variations are shown in Figure \ref{fig:Ex2_rand_bias_init_recoveries} (a) and range from about -0.3 to about 0.3 km/s. The errors in the macro velocity model of the target area have the largest effect on the data at the bottom and the deeper parts of the side boundaries. Therefore we use the biased true model to generate the data at the bottom and side boundaries, while the top boundary data is still exact. We invert this data using the biased initial model.

\begin{figure}\centering
 	\includegraphics[width = 0.8\textwidth,trim={0cm 0cm 0cm 0cm},clip]{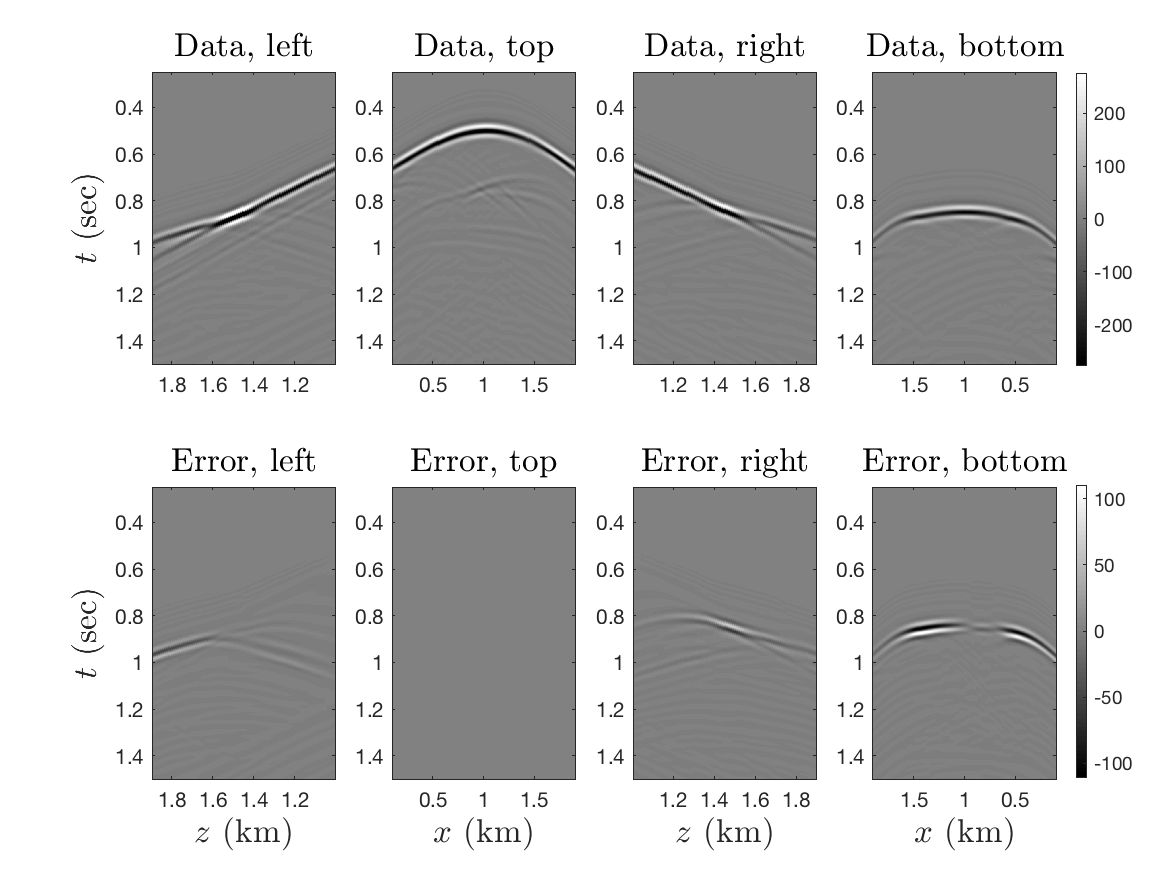}
	\caption{(\textit{top row}) Pressure data on the injection boundary, where the bottom, left and right boundary data is obtained from the kinematically biased true model; (\textit{bottom row}) Data error. }
 	\label{fig:Ex2_data_error}
\end{figure}

Figure \ref{fig:Ex2_data_error} shows the pressure field on the boundary used for this example and pressure data error. As expected, the error is the largest at the bottom boundary and increases with depth at the side boundaries. Figure \ref{fig:Ex2_rand_bias_init_recoveries} (b) shows the IFWI reconstruction after 109 iterations. After this the image is still reasonable but becomes slightly grainy. We observe that the inversion is able to reproduce most of the high wavenumber details in the model, although some of the low wavenumber bias propagates into the reconstruction, particularly, the low velocity in the right and left bottom corners. 

While some insight can be gleaned from this proxy example, a more accurate assessment needs to be made with redatumed data.

\subsubsection{Example 4: missing data} \label{missing_data}

\begin{figure}\centering
 	\subfigure[]{\includegraphics[width = 0.45\textwidth,trim={0cm 1.5cm 0cm 3cm},clip]{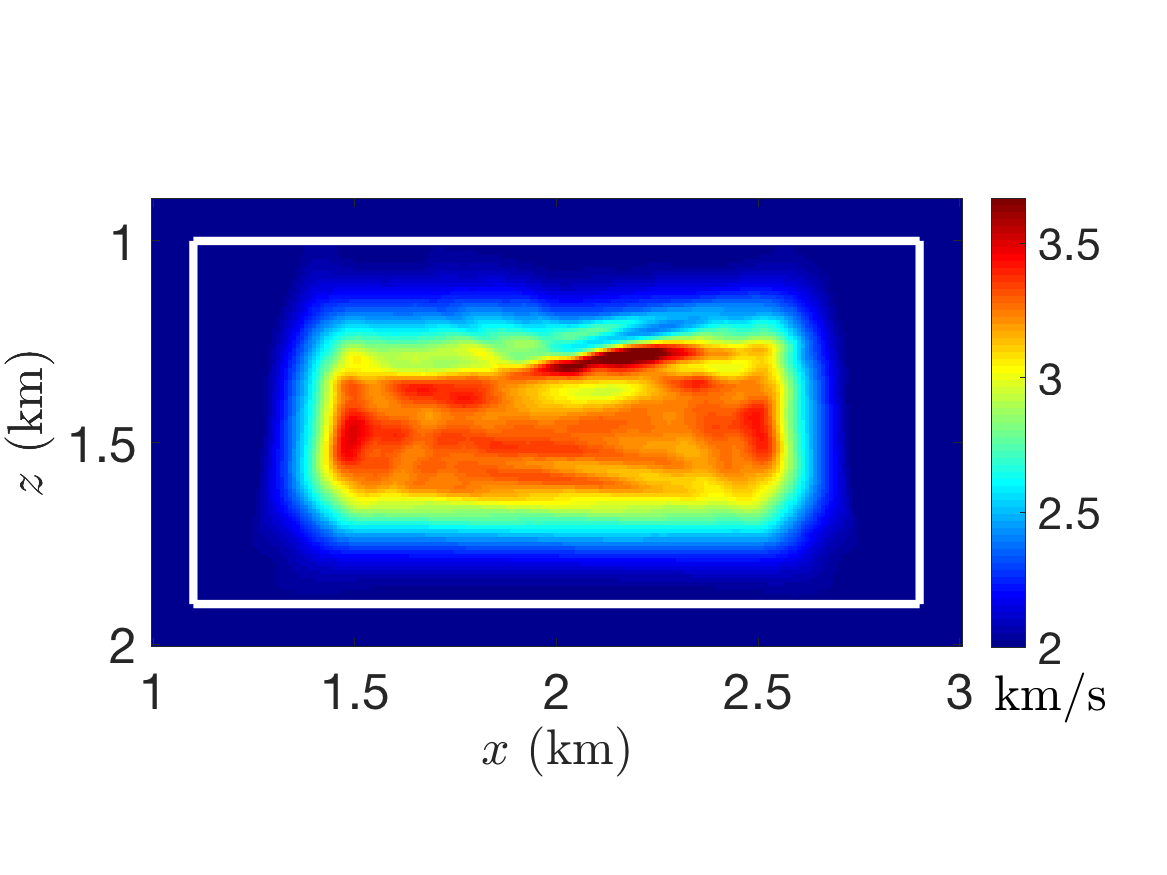}} 
 	\subfigure[]{\includegraphics[width = 0.45\textwidth,trim={0cm 1.5cm 0cm 3cm},clip]{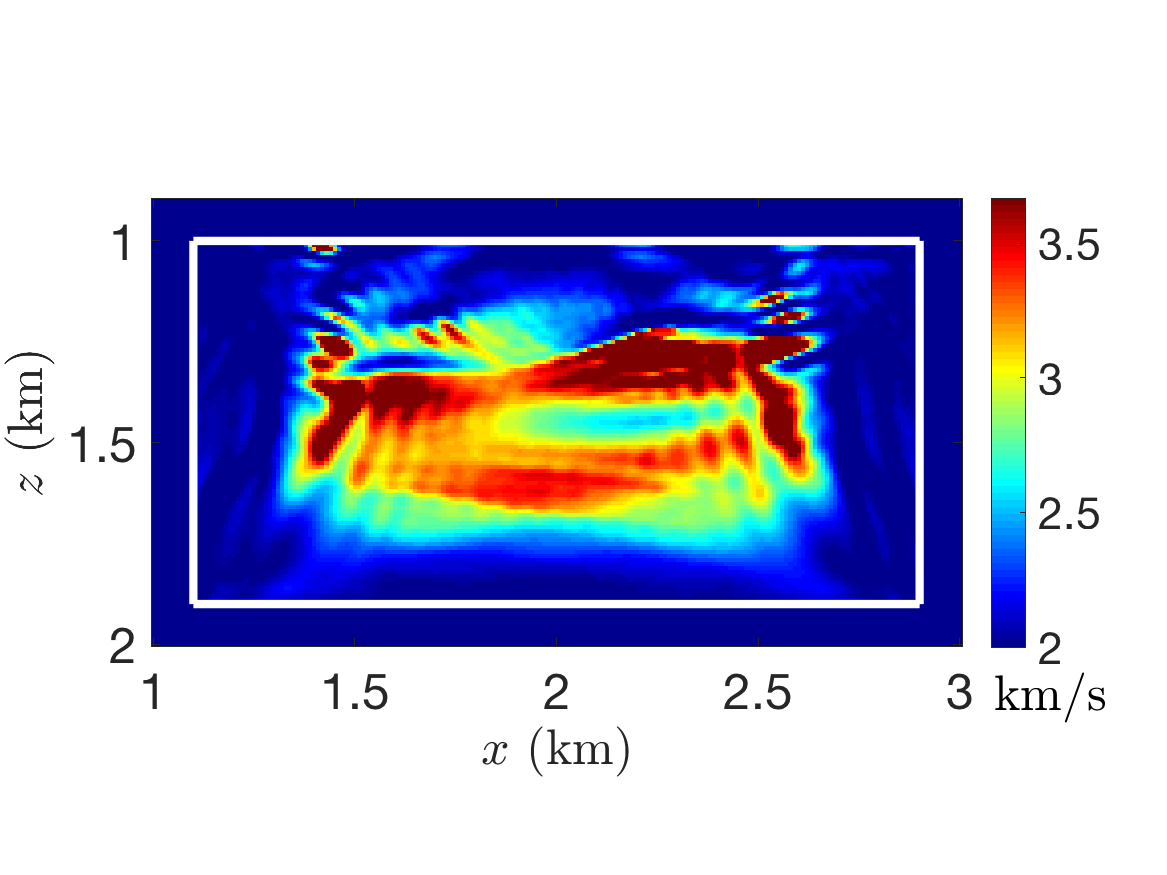}} 
	\caption{IFWI reconstruction with the top and bottom boundaries only after (a) 5 and (b) 41 iterations.}
 	\label{fig:Ex3_nosides}
\end{figure}

In this section, we show the effect of missing side boundary data on the inversion. To this end, we perform IFWI with top and bottom boundaries only. Figures \ref{fig:Ex3_nosides} (a) and (b) show the inversion progress at iteration 5 and the final inverted model at iteration 41. The inversion stopped after 41 iterations due to a failed line search. We observe that the absence of the data on the side boundaries introduces artifacts in the form of side boundary reflections that become larger with iteration and eventually cause the inversion to fail. This behaviour is expected because the convolution and correlation representation theorems are only valid in the presence of an enclosing boundary or infinite top and bottom boundaries.

\section{Discussion}

At this point in the development of our new waveform inversion algorithm, there are still a number open questions that require further investigation. More specifically:
\begin{enumerate}
    \item Robustness to inaccurate redatumed fields: although our results are currently based on exact boundary wavefields, both \cite{Costa:2018} and \cite{Cui:2020} have shown that model-based redatuming algorithms such as full wavefield migration and data-driven approaches such as Marchenko redatuming, can produce wavefields of satisfactory quality for convolution-based local inversion. Whilst our cost function relies also on a representation theorem of correlation type, as well as the inclusion of side and bottom boundaries, we expect a similar robustness to small errors in the boundary wavefields. However, due to one-sidedness and limited aperture of surface seismic data acquisition, data-driven approaches such as Marchenko redatuming are likely to best retrieve waves propagating near-vertically. Therefore, we can expect redatumed data on the side boundaries to be less accurate than the data on the top and bottom boundaries. Specifically, \citep{Cui:2020} speculate that the redatumed pressures on the side boundaries might not provide an accurate enough horizontal displacement component necessary for the application of the representation formulas. It is possible, however, that in light of recent revisions of Marchenko formalism \citep{Diekmann:2021, Wapenaar:2021, Kiraz:2021} future versions of Marchenko-based redatuming may provide improved estimates of wavefields on vertical/side boundaries. In any case, it is likely that large horizontal offsets might be necessary for more accurate redatuming of the data on the side boundaries and close to the top and bottom corners. The example in section \ref{missing_data} also shows that complete exclusion of the horizontal boundaries has a devastating effect on the inversion. In theory, it is only possible to exclude the side boundaries if the local domain has infinite width. In practice, extending the local domain in width also leads to an increase in the computational cost and the number of model parameters in the inversion with unclear effect on the inversion. Assessment of the effect of data redatuming, particularly at the side boundaries, on the inversion is the subject of our current research work. 
    
    \item Low wavenumber updates: whilst the ability of the interferometric objective function to retrieve low wavenumber components in a target area is quite notable, this heavily depends on kinematic features captured by transmissions in the data (mostly along the bottom boundary). Therefore, the applicability of the proposed method for the macro model update may be limited for two reasons: (i) an accurate macro velocity model is still required to obtain kinematically correct redatumed data at the boundaries, and (ii) the kinematic errors in the boundary data may well  lead to incorrect retrieval of the low wavenumber components in the model by inversion. The proxy example in section \ref{incorrect data} shows that if kinematic errors from the target area are present in the data at the side and bottom boundaries, they propagate into the reconstruction. Therefore, in its current implementation, the proposed method is more suitable to localized model refinement rather than to macro velocity model building. An alternative update strategy may be necessary to mitigate the incorrect low wavenumber components in the redatumed data. Furthermore, more research is needed to assess the influence of redatumed-waveform errors caused by the overburden versus those related to the target medium, as we expect these to affect our method in different ways.
    
   \item Distribution and density of volume points: the proposed objective function can be evaluated at any point inside the enclosing boundary. This plays the counterpart of physical receivers in conventional and local FWI. However, whilst the distribution and number of receivers is fixed and dictated by physical (in FWI) or algorithmic (in local FWI) constraints, the choice of the grid of points where the cost function is evaluated is totally arbitrary in our case. Our current implementation relies on a dense grid where the spatial sampling equates to that of the FD grid used for modelling of the wavefields. The use of coarser (or finer), and possibly iteration dependent grids e.g., to balance bandwidth-related resolution with memory usage), will be a subject of future studies. For example, we envisage the combining low-frequency, coarse-grid FWI for redatuming-ready background and overburden models, with wide-bandwidth, denser-grid IFWI target inversion.
       
    \item Cost function evaluation outside of the enclosing boundary: an additional feature of the proposed objective function is that it can also be evaluated outside of the enclosing boundary. In this case, it is not only the difference between the convolution and correlation representation theorems that will be zero in the presence of a correct model, but also each individual term alone must vanish as a result of writing the convolution and correlation representation formulas when both sources are outside of the boundary. This property is however not satisfied when the Green's function $\Gb_D$ is modelled in the incorrect medium. Future research will investigate the benefit of enriching the interferometric objective function with grid points outside of the boundary.
    
    \item Multi-parameter and elastic inversion: whilst the numerical example presented in this paper targets only one parameter (i.e., velocity), the proposed framework is suited to multi-parameter inversion, namely (visco)elasticity and density, in that it relies on representation theorems that contain and reconstruct both pressure and particle velocity recordings -- or to include particle velocity and stress fields in the more general cases. Moreover, extension to elastic media is also straightforward using the (visco)elastic counterparts of the convolution and correlation representation theorems  \citep{Aki:2002} -- with the inclusion of viscoelasticity being also possible, granted the representation terms then include the appropriate volume terms. Moreover, given the much-increased computational costs associated with elastic media, our target-oriented approach may prove an essential tool to achieve detailed inverted models of target reservoirs at depth.
\end{enumerate}

\section{Conclusions}

We propose a novel, target-enclosing, waveform inversion method based on an interferometric objective function that minimizes the difference between wavefields reconstructed from the boundary data by convolution and correlation representation formulas. The representation formulas are reformulated as PDEs constraints for the inversion. The method is formulated using full vector-acoustic boundary data consisting of pressure and scaled particle displacement. We derive the gradient of the objective with respect to the model parameter and implement the inversion using L-BFGS optimization engine. The method is shown to be very suitable for high-resolution model refinement, provided that an accurate macro velocity model is available.

In this initial work, we test the new inversion algorithm on stylized examples. We find that the objective function is also able to recover the low wavenumbers in the model better than other waveform inversion techniques when the boundary data is exact, i.e. contains the correct information about the macro velocity model. This seems to affect the resolution. In practice, the quality of the reconstruction also depends on the accuracy of redatuming, which in turn relies on the kinematic velocity model. Therefore, the ability of IFWI to recover the long wavenumber components is likely to be limited in practice. From the numerical examples, a key feature of the method appears to be evaluation of the objective function everywhere in the local subdomain. This is achieved at the expense of two extra local domain PDE solves per iteration, compared to the closest competitor method: the local convolution-based FWI method of \citep{Cui:2020}. Considering that comparable resolution for the local FWI method is likely to require extra redatuming effort, our proposed method appears to be at least competitive in cost.

Further assessment of the robustness of the proposed technique with respect to redatuming is necessary. Future work will be focused on combining the proposed objective function with boundary data redatumed by means of the Marchenko method both on synthetic and field data.

\section{Acknowledgments}
The authors thank King Abdullah University of Science and Technology (KAUST) for funding this work. For computer time, this research used the resources of the Supercomputing Laboratory at King Abdullah University of Science \& Technology (KAUST) in Thuwal, Saudi Arabia. Alison Malcolm thanks NSERC, Chevron and InnovateNL for funding.

\section{Data Availability}

The codes for this manuscript can be made available upon request.

\appendix
\section{IFWI gradient derivation} \label{AppA}
We begin by defining the convolution and correlation forward modelling operators:
\begin{eqnarray}
    \Fm^{conv}(m): m \in \Mm \mapsto \wb^{conv} \in \Dm\\
    \Fm^{corr}(m): m \in \Mm \mapsto \wb^{corr} \in \Dm 
\end{eqnarray}
where $\Mm$ and $\Dm$ are the model and data spaces respectively, and the mapping is computed by solving equations (\ref{conv_constr}) and (\ref{corr_constr}) .

The inner products in the model and data spaces are defined as follows:
\begin{eqnarray}
    \lan m_1 , m_2 \ran_{\Mm} &=& \int_D dV [m_1(\x) m_2(\x)] \label{model_ip}\\
    \lan \wb_1 , \wb_2 \ran_{\Dm} &=& \int_0^T  dt \int_D dV [\wb_1(\x,t) \cdot  \wb_2(\x,t)], \label{data_ip}
\end{eqnarray}
which makes $\Mm$ and $\Dm$ Hilbert spaces with the corresponding induced norms, so that the objective function in equation \ref{objective} can be rewritten as:
\begin{equation}
	I(m) = \lan \Wr [ \Fm^{corr}(m) - \Fm^{conv}(m)], \Wr [ \Fm^{corr}(m) - \Fm^{conv}(m)] \ran_{\Dm} \label{obj_as_inner_prod}
\end{equation}

We also define the linearized forward maps 
\begin{eqnarray}
	\Fb^{conv}[m]: \delta m \in \Mm \mapsto \delta \wb^{conv} \in \Dm\\
	\Fb^{corr}[m]: \delta m \in \Mm \mapsto \delta \wb^{corr} \in \Dm
\end{eqnarray}
that are evaluated by solving the linearized forward modelling equations:
\begin{eqnarray}
    \Lb \delta \wb^{conv} = -\delta m \Wr \wb^{conv} \label{lin_conv}\\
    \Lbdag \delta \wb^{corr} = -\delta m \Wr \wb^{corr} \label{lin_corr},
\end{eqnarray}
where equations (\ref{lin_conv}) and (\ref{lin_corr}) are solved forward and backward in time respectively.
Then, one can show that
\begin{eqnarray}
    \Fm^{conv}(m+\delta m) = \Fm^{conv}(m) + \Fb^{conv}[m]\delta m + O(\|\delta m\|^2) \label{conv_exp}\\
    \Fm^{corr}(m+\delta m) = \Fm^{corr}(m) + \Fb^{corr}[m]\delta m + O(\|\delta m\|^2). \label{corr_exp}
\end{eqnarray}

We consider the difference $I(m + \delta m) - I(m)$. Using (\ref{obj_as_inner_prod}), (\ref{conv_exp}) and (\ref{corr_exp}) we can show that
\begin{equation} \begin{split}
    &I(m + \delta m) - I(m) = \\
    & + \lan \Wr (\Fm^{corr}(m) - \Fm^{conv}(m)),  \Wr ( \Fb^{corr}[m]\delta m - \Fb^{conv}[m] \delta m) \ran_{\Dm} + \\
    &+ O(\|\delta m\|^2) = \\
    &= \lan \Wradj \Wr (\Fm^{corr}(m) - \Fm^{conv}(m)),  \delta \wb^{corr} \ran_{\Dm} - \\
    &- \lan \Wradj \Wr (\Fm^{corr}(m) - \Fm^{conv}(m)),  \delta \wb^{conv} \ran_{\Dm} + \\
    &+ O(\|\delta m\|^2)
    \label{linearization}
\end{split} \end{equation}

We introduce the adjoint problem (\ref{adj_conv}) and (\ref{adj_corr})
\begin{eqnarray}
    \Lbdag \wb^{conv \dagger} = - \Wradj \Wr (\Fm^{corr}(m) - \Fm^{conv}(m)) \nonumber \\
    \Lb \wb^{corr \dagger} = \Wradj \Wr (\Fm^{corr}(m) - \Fm^{conv}(m)) \nonumber,
\end{eqnarray}
where equations (\ref{adj_corr}) and (\ref{adj_conv}) are solved respectively forward and backward in time. 
Then, with the help of (\ref{lin_conv}) and (\ref{lin_corr}), equation (\ref{linearization}) becomes:
\begin{equation} \begin{split}
    &I(m + \delta m) - I(m) = \\
    &= \lan \Lb \wb^{corr \dagger},  \delta \wb^{corr} \ran_{\Dm} + \lan \Lbdag \wb^{conv \dagger},  \delta \wb^{conv} \ran_{\Dm} + O(\|\delta m\|^2) = \\
    & = \lan \wb^{corr \dagger}, \Lb^{\dagger} \delta \wb^{corr} \ran_{\Dm} + \lan \wb^{conv \dagger}, \Lb \delta \wb^{conv} \ran_{\Dm}  + O(\|\delta m\|^2) = \\
    & = \lan \wb^{corr \dagger}, - \Wr \wb^{corr} \delta m \ran_{\Dm}  + \lan \wb^{conv \dagger}, - \Wr \wb^{conv} \delta m \ran_{\Dm} + O(\|\delta m\|^2) = \\
    & = \lan \int_0^T(-\wb^{corr,T} \Wr \wb^{corr \dagger}) dt, \delta m \ran_{\Mm} +\\
    &\lan \int_0^T (- \wb^{conv, T}  \Wr \wb^{conv \dagger}) dt,  \delta m \ran_{\Mm} + O(\|\delta m\|^2) = \\
    & = \lan \int_0^T (-p^{corr} p^{corr \dagger} - p^{conv} p^{conv \dagger}) dt, \delta m \ran_{\Mm} + O(\|\delta m\|^2) 
    \label{linearization2}
\end{split} \end{equation}
By definition of Frechet derivative, we  conclude from (\ref{linearization2}) that
\begin{equation}
	\frac{\partial I}{\partial m} = \int_0^T (-p^{corr} p^{corr \dagger} - p^{conv} p^{conv \dagger}) dt.
\end{equation}


\label{lastpage}

\end{document}